\documentclass[aps,prd,superscriptaddress,nofootinbib,amsmath,amsfonts,preprintnumbers,groupedaddress,showpacs,%dvipdfmx,
10pt,english]{revtex4-1}
\usepackage{amsmath}
\usepackage{amssymb}
\usepackage{babel}
\usepackage{wrapfig}
\usepackage{cancel}

\newcommand{\e}{\mathrm{e}}
\usepackage{relsize,exscale}
\makeatletter

\usepackage{array,multirow,graphicx}
\usepackage{dcolumn}
\usepackage{newlfont}
\usepackage{bm}
\usepackage[colorlinks,citecolor=blue,urlcolor=blue,linkcolor=blue]{hyperref}
\usepackage[figtopcap]{subfigure}
\usepackage{color}

\allowdisplaybreaks[4]

\begin{document}
\tolerance=5000

\title{Multi-horizons black hole solutions, photon sphere and perihelion shift \\
in weak ghost-free Gauss-Bonnet theory of gravity}
\author{G.~G.~L.~Nashed}
\email{nashed@bue.edu.eg}
\affiliation {Centre for Theoretical Physics, The British University in Egypt, P.O. Box
43, El Sherouk City, Cairo 11837, Egypt}
\author{Shin'ichi~Nojiri}
\email{nojiri@gravity.phys.nagoya-u.ac.jp}
\affiliation{Department of Physics, Nagoya University, Nagoya 464-8602,
Japan \\
\& \\
Kobayashi-Maskawa Institute for the Origin of Particles and the Universe,
Nagoya University, Nagoya 464-8602, Japan }

%\date{\today}

\begin{abstract}
Among the modified gravitational theories, the ghost-free Gauss-Bonnet (GFGB) theory of gravity has been considered from the viewpoint of cosmology.
The best way to check its applicability could be to elicit observable predicts which give guidelines or limitations on the theory, which could be contrasted with the actual observations.
In the present study, we derive consistent field equations for GFGB and by applying the equations to a spherically symmetric space-time, we
obtain new spherically symmetric black hole (BH) solutions.
We study the physical properties of these BH solutions and show that the obtained space-time possesses multi-horizons and the Gauss-Bonnet invariants in the space-time are not trivial.
We also investigate the thermodynamical quantities related to these BH solutions and we show that these quantities are consistent with what is known in the previous works.
Finally, we study the geodesic equations of these solutions which give the photon spheres and we find the perihelion shift for weak GFGB.
In addition, we calculate the first-order GFGB perturbations in the Schwarzschild solution and new BH solutions and show that we improve and extend existing results
in the past literature on the spherically symmetric solutions.

\end{abstract}

\maketitle

\section{Introduction}

Although more than one hundred years have passed after the construction of Einstein's theory of general relativity (GR),
GR is still the most established macroscopic theory of gravity that is widely accepted.
In spite of its vast success in both weak and strong couplings \cite{Will:2014kxa,Ishak:2018his}, however, there is still no harmonic way to link the macroscopic theory of GR
to a quantum field theory.
%Regardless of this, GR still has no satisfactory answer to the issue of local energy-momentum conservation.
Moreover, GR predicts space-time singularity which has mathematical results in its construction.
The problem of singularity leads scientists to search for other theories of gravity that could coincide with GR in the scale of daily life and/or the scale of the solar system.
It is interesting to note that Lovelock's theory~\cite{Lovelock:1971yv} has explained that in four dimensions, Einsteins GR is the unique metric theory of gravity
that could yields symmetric, covariant second-order field equations.
Therefore, one of the attempts to amend Einstein's GR is to work in space-times with extra dimensions \cite{Clifton:2011jh}.
In these attempts, the most general set of theories could be the Lovelock theories which yield symmetric, covariant second-order field equations
regarding the metric tensor in any space-time dimensions \cite{Padmanabhan:2013xyr}.
The Lagrangian of the Lovelock theory is given as follows,
\begin{align}
\mathcal{L}=\sqrt{-g}\left(-2 \Lambda + R + \alpha \mathcal{G}+\cdots\right)\, ,
\end{align}
with $\mathcal{G}\equiv R^2 - 4 R_{\mu \nu} R^{\mu \nu}+ R_{\alpha \beta \mu \nu}R^{\alpha \beta \mu \nu}$ being the Gauss-Bonnet (GB) invariant
which yields the first order correction to the action of Einstein's theory with a cosmological constant $\Lambda$.
Although the GB invariant yields nontrivial effects when the space-time dimensions are larger than four, the invariant is topological in four dimensions \cite{10.2307/1969203}.
Regardless of being quadratic in curvatures, the GB invariant has theoretical wide advantages from the viewpoints of string
theory \cite{Ferrara:1996hh, Antoniadis:1997eg, Zwiebach:1985uq, Nepomechie:1985us, Callan:1986jb, Candelas:1985en,Gross:1986mw}.

Many researchers have been tempted by the idea of harmony merging the effect of the GB invariant in a four-dimensional theory of gravity,
which could yield equations of motions different from GR, avoiding Lovelock's theorem.
Glavan and Lin~\cite{Glavan:2019inb} have investigated the idea to rescale the GB coupling constant $\gamma$ in $N$ dimensions as
$\gamma \rightarrow \gamma/(N-4)$, so that there remains the contribution from the GB invariant in the limit $N \to 4$.
After that, there have appeared works, where spherical black hole solutions\cite{Kumar:2020uyz,Fernandes:2020rpa,Kumar:2020owy,Ghosh:2020syx},
the construction of cosmological solutions \cite{Li:2020tlo, Kobayashi:2020wqy},
the radiation of black holes and the collapse to the black hole \cite{Ghosh:2020vpc,Shirafuji:1996im,Malafarina:2020pvl},
star-like objects \cite{Doneva:2020ped}, the extension to more higher-curvature Lovelock theories \cite{Konoplya:2020qqh},
the thermodynamical behavior of black hole solutions \cite{EslamPanah:2020hoj, HosseiniMansoori:2020yfj,Konoplya:2020cbv,Hegde:2020xlv},
and the physical properties of such objects \cite{Guo:2020zmf,Konoplya:2020qqh,Zhang:2020qew, Roy:2020dyy, NaveenaKumara:2020kpz,Liu:2020vkh,
Heydari-Fard:2021ljh, Kumar:2020sag, Nashed:2018efg,Islam:2020xmy,Mishra:2020gce,Devi:2020uac,Churilova:2020mif} have been investigated .
In spite of all of these researches, the regularization method used in the four-dimensional Einstein-GB theory \cite{Glavan:2019inb}
has been shown to be inconsistent for many reasons \cite{Gurses:2020ofy,Gurses:2020rxb,Arrechea:2020evj, Arrechea:2020gjw,Bonifacio:2020vbk,Nashed:2018cth,Ai:2020peo,
Mahapatra:2020rds,Nashed:2010ocg,Hohmann:2020cor,Cao:2021nng}, which yield to the construction of different models of the regularized (harmonic)
four-dimensional Einstein GB theories~\cite{Lu:2020iav,Kobayashi:2020wqy,Fernandes:2020nbq, Hennigar:2020lsl, Aoki:2020lig, Fernandes:2021dsb}.

Due to several reasons, some researchers are suspicious about the procedure proposed in \cite{Glavan:2019inb}.
One reason is that the field equations of the Einstein-GB theory defined in higher dimensions can be divided into two various sets.
One set yields the field equations which always come from higher dimensional theories and this set makes the specific action in the limit of $N\to4$
non-trivial \cite{Gurses:2020ofy, Gurses:2020rxb, Arrechea:2020evj, Nashed:2007cu,Arrechea:2020gjw,Mahapatra:2020rds}.
The tree-level graviton scattering amplitude was also investigated in this frame, apart from the Lagrangian,
and it turned out that the dimensional continuation, $N\to 4$, does not make the GB amplitude create any new four-dimensional GB gravitational
amplitude \cite{Lin:2020kqe}.
All of these attempts yield the fact that the existence of the solution in the limit of $N\to 4$ does not mean that there is a four-dimensional theory
as proposed in \cite{Glavan:2019inb}.
In spite of this situation, it could be important to mention that the field equations different from the four-dimensional Einstein GB
gravity~\cite{Lu:2020iav,Kobayashi:2020wqy, Fernandes:2020nbq,Hennigar:2020lsl, Aoki:2020lig} support the same static spherically symmetric BH solution
as constructed in \cite{Glavan:2019inb}.
Following the $N\to 4$ regularization of the scalar and vector type gravitational perturbation of the $N>4$ Einstein-GB BH~\cite{Takahashi:2010gz, Takahashi:2010ye},
it has been investigated that the asymptotically flat or AdS/dS BHs are unstable for large positive values of
the GB coupling parameter~\cite{Konoplya:2020bxa, Konoplya:2020juj}.
The quasinormal modes of the four-dimensional Einstein GB BH in the asymptotically AdS/dS space-time due to scalar, electromagnetic, and Dirac perturbations
have been investigated in~\cite{Devi:2020uac, Churilova:2020mif}.
The quasi-bound states of massless scalar, electromagnetic, and Dirac fields in the asymptotically flat four-dimensional Einstein GB BH
and the associated stability issue have been studied in \cite{Vieira:2021doo}.

Because of the significance of the theories involving the GB scalar, which are encouraged by string theories in many cases,
in this study, we shall briefly discuss the drawback of these theories, specifically the existence of ghosts.
Generally, higher-derivative gravitational theories involve ghost degrees of freedom due to the Ostrogradskya's instability \cite{Woodard:2015zca}.
It was explained in \cite{DeFelice:2009ak}, that ghost degrees of freedom could happen at different levels of
the theory, despite the cosmological perturbations level of $f(R, \mathcal{G})$ theories.
It is the aim of the present study to derive the spherically symmetric BH solutions in the ghost-free $f\left(\mathcal{G} \right)$ gravitational theory
proposed in \cite{Nojiri:2018ouv, Nojiri:2021mxf}.

%%%%%%%%%%%%%%%%
%%%%%%%%%%%%

This paper is organized as follows:
In Section~\ref{S2}, we present the basic tools for the ghost-free $f\left(\mathcal{G} \right)$ gravitational theory that is capable to describe the formulation of BH horizons.
In Section~\ref{S3}, we apply the field equations of GFGB to the spherically symmetric space-time and derive BH solutions with multi-horizons.
In Section~\ref{S4}, we study the relevant physics of the BH solutions derived in Section~\ref{S3} by showing their asymptote at $r\to \infty$.
Moreover, we show that by studying the thermodynamical behavior of these BH solutions by calculating their thermodynamical quantities like Hawking temperature,
heat capacity, and the Gibbs free energy, we show that all these quantities related to the BHs derived in Section~\ref{S2} are consistent with
the results presented in the past literature.
In Section~\ref{sec:pheno}, we study the particle motion phenomenology for these BHs and derived their potential for the Schwarzschild background.
Moreover, we derive the deviation from Einstein's general relativity of the photon sphere and the perihelion shift.
We close our study with the conclusion of the main results in Section~\ref{con}.

Throughout the present study, we assume the relativistic units, i.e., $G=c=1$.

\section{Brief summary of ghost-free $f(\mathcal{G})$ gravitational theory}\label{S2}

In the present section, we will present briefly the ghost-free$f\left(\mathcal{G} \right)$ gravity in the formulation using the Lagrange multipliers.
Moreover, we shall investigate how to obtain a ghost-free $f(\mathcal{G})$ gravity, and we shall employ the Lagrange multipliers formalism in order to achieve this.
Before going to the details of the formalism, we will start the derivation by showing in detail how ghost modes could exist in $f\left(
\mathcal{G} \right)$ gravity at the field equations level, and then construct
the ghost-free model construction of the theory.

\subsection{Ghosts in $f\left( \mathcal{G} \right)$ Gravity}

Nojiri et al. \cite{Nojiri:2018ouv, Nojiri:2021mxf} have constructed a ghost-free $f\left(\mathcal{G} \right)$ gravity theory by using the Lagrange multiplier field.
The original $f\left(\mathcal{G} \right)$, whose action to has the following form,
\begin{align}
\label{FRGBg}
S=\int d^4x\sqrt{-g} \left(\frac{1}{2\kappa^2}R
+ f\left( \mathcal{G}\right) + \mathcal{L}_\mathrm{matter}\right)\, ,
\end{align}
have ghost as we show below.
Here $\mathcal{L}_\mathrm{matter}$ is the Lagrangian density of the matters.
The above action (\ref{FRGBg}) can be rewritten as follows,
\begin{align}
\label{FRGBg12}
S=\int d^4x\sqrt{-g} \left(\frac{1}{2\kappa^2}R
+ h\left( \chi \right) \mathcal{G} - V\left( \chi \right)
+ \mathcal{L}_\mathrm{matter}\right)\, ,
\end{align}
where $R$ is the Ricci scalar, $\chi$ is an auxiliary field, $\mathcal{G}$ is the GB invariant, $V(\chi)$ is the potential and $h(\chi)$
is a function of the auxiliary field.
The variation of the action (\ref{FRGBg12}) w.r.t. the $\chi$, gives,
\begin{align}
\label{FRGBg13}
0 = h'\left( \chi \right) \mathcal{G} - V'\left( \chi \right) \, .
\end{align}
Eq.~(\ref{FRGBg13}) can be solved w.r.t. $\chi$ as a function of the GB invariant $\mathcal{G}$,
$\chi = \chi \left( \mathcal{G} \right)$.
Then by substituting the obtained expression of $\chi\left( \mathcal{G} \right)$ into Eq.~(\ref{FRGBg13}), one can reobtain the action of
Eq.~(\ref{FRGBg}) where $f \left( \mathcal{G} \right)$ is defined as,
\begin{align}
\label{FRGBg14}
f \left( \mathcal{G} \right) = h \left( \chi \left( \mathcal{G} \right) \right) \mathcal{G}
 - V \left( \chi \left( \mathcal{G} \right) \right) \, .
\end{align}
Furthermore, the varying of the action (\ref{FRGBg13}) w.r.t. the metric tensor yields:
\begin{align}
\label{FRGBg15}
0 = \frac{1}{2\kappa^2}\left(- R_{\mu\nu}
+ \frac{1}{2}g_{\mu\nu} R\right) + \frac{1}{2} T_{\mathrm{matter}\, \mu\nu}
 - \frac{1}{2}g_{\mu\nu} V \left( \chi \right)
+ D_{\mu\nu}^{\ \ \tau\eta} \nabla_\tau \nabla_\eta h \left( \chi
\right) \, ,
\end{align}
where the tensor $D_{\mu\nu}^{\ \ \tau\eta}$ is defined:
\begin{align}
\label{fRGB7}
 D_{\mu\nu}^{\ \ \tau\eta} \equiv&\, \left( \delta_\mu^{\
\tau} \delta_\nu^{\ \eta} + \delta_\nu^{\ \tau} \delta_\mu^{\
\eta} - 2 g_{\mu\nu} g^{\tau\eta} \right) R + \left( - 4
g^{\rho\tau} \delta_\mu^{\ \eta} \delta_\nu^{\ \sigma}
 - 4 g^{\rho\tau} \delta_\nu^{\ \eta} \delta_\mu^{\ \sigma}
+ 4 g_{\mu\nu} g^{\rho\tau} g^{\sigma\eta} \right) R_{\rho\sigma} \nonumber \\
&\, + 4 R_{\mu\nu} g^{\tau\eta}
 - 2 R_{\rho\mu\sigma \nu} \left( g^{\rho\tau} g^{\sigma\eta}
+ g^{\rho\eta} g^{\sigma\tau} \right) \, .
\end{align}
Since the auxiliary field $\chi$ can be rewritten as a function of the
GB $\mathcal{G}$, then Eq.~(\ref{FRGBg15}) is a
fourth order differential equation for the metric which may contain ghost modes.

In order to eliminate the ghost modes, we may add a canonical kinetic term of $\chi$ in the action (\ref{FRGBg12})
\begin{align}
\label{FRGBg16}
S=\int d^4x\sqrt{-g} \left(\frac{1}{2\kappa^2}R
+ h\left( \chi \right) \mathcal{G}
 - \frac{1}{2} \partial_\mu \chi \partial^\mu \chi
- V\left( \chi \right) +
\mathcal{L}_\mathrm{matter}\right)\, ,
\end{align}
where we have chosen the mass dimension of $\chi$ to be unity.
Then variation of the action (\ref{FRGBg16}) w.r.t. $\chi$ and metric give \cite{Nojiri:2005vv,Nojiri:2018ouv, Nojiri:2021mxf},
\begin{align}
\label{FRGBg17}
0 =& \Box \chi + h'\left( \chi \right) \mathcal{G} - V'\left( \chi \right) \, , \\
\label{FRGBg18}
0 =& \frac{1}{2\kappa^2}\left(- R_{\mu\nu}
+ \frac{1}{2}g_{\mu\nu} R\right) + \frac{1}{2} T_{\mathrm{matter}\, \mu\nu}
+ \frac{1}{2} \partial_\mu \chi \partial_\nu \chi
 - \frac{1}{2}g_{\mu\nu} \left( \frac{1}{2} \partial_\rho \chi \partial^\rho \chi
+ V \left( \chi \right) \right)
+ D_{\mu\nu}^{\ \ \tau\eta} \nabla_\tau \nabla_\eta h \left( \chi \right) \, .
\end{align}
The equations derived in Eq.~(\ref{FRGBg17}) do not have higher-order except the
second-order derivatives which mean that we could not have ghosts.

The model (\ref{FRGBg16}), has a new dynamical degree of freedom, i.e., $\chi$, however, if we like to
minimize the dynamical degrees of freedom, we can insert a
constraint as in the mimetic theory~\cite{Chamseddine:2013kea, Nojiri:2014zqa,Dutta:2017fjw}, by
using the Lagrange multiplier field $\lambda$, as follows,
\begin{align}
\label{FRGBg19}
S=\int d^4x\sqrt{-g} \left(\frac{1}{2\kappa^2}R
+ \lambda \left( \frac{1}{2} \partial_\mu \chi \partial^\mu \chi + \frac{\mu^4}{2} \right)
 - \frac{1}{2} \partial_\mu \chi \partial^\mu \chi
+ h\left( \chi \right) \mathcal{G} - V\left( \chi \right) +
\mathcal{L}_\mathrm{matter}\right)\, ,
\end{align}
where $\mu$ is a constant which has a mass dimension.
Thus, by varying action (\ref{FRGBg19}) w.r.t. $\lambda$, we obtain,
\begin{align}
\label{FRGBg20}
0=\frac{1}{2} \partial_\mu \chi \partial^\mu \chi + \frac{\mu^4}{2} \, .
\end{align}
Because of the fact that the kinetic term becomes a constant, the kinetic term in Eq.~(\ref{FRGBg19})
can be absorbed by using the redefinition of potential $V\left( \chi\right)$,
\begin{align}
\label{FRGBg21}
\tilde V \left(\chi\right) \equiv \frac{1}{2} \partial_\mu \chi \partial^\mu \chi
+ V \left( \chi \right) = - \frac{\mu^4}{2} + V \left( \chi \right) \,.
\end{align}
%\newpage
Now, the action of Eq.~(\ref{FRGBg19}) can be rewritten as follows,
\begin{align}
\label{FRGBg22}
S=\int d^4x\sqrt{-g} \left(\frac{1}{2\kappa^2}R
+ \lambda \left( \frac{1}{2} \partial_\mu \chi \partial^\mu \chi + \frac{\mu^4}{2} \right)
+ h\left( \chi \right) \mathcal{G} - \tilde V\left( \chi \right)
+ \mathcal{L}_\mathrm{matter}\right)\, .
\end{align}
The action given in Eq.~(\ref{FRGBg22}) yields, in addition to
Eq.~(\ref{FRGBg20}), the following two equations of motion,
\begin{align}
\label{FRGBg23_0}
% 0=&\frac{1}{2}\omega(\chi) \partial_\mu \chi \partial^\mu \chi + \frac{\mu^4}{2}\,,\\
% \label{FRGBg23gg}
0 =& - \frac{1}{\sqrt{-g}} \partial_\mu \left( \lambda \omega(\chi) g^{\mu\nu}\sqrt{-g}
\partial_\nu \chi \right)
+ h'\left( \chi \right) \mathcal{G} - {\tilde V}'\left( \chi \right) +\frac{1}{2} \lambda \omega'(\chi) g^{\mu\nu}
\partial_\mu \chi\partial_\nu \chi\, ,\\
\label{FRGBg24}
0 =& \frac{1}{2\kappa^2}\left(- R_{\mu\nu}
+ \frac{1}{2}g_{\mu\nu} R\right) + \frac{1}{2} T_{\mathrm{matter}\, \mu\nu}
 - \frac{1}{2} \lambda \partial_\mu \chi \partial_\nu \chi
 - \frac{1}{2}g_{\mu\nu} \tilde V \left( \chi \right)
+ D_{\mu\nu}^{\ \ \tau\eta} \nabla_\tau \nabla_\eta h \left( \chi
\right)\, .
\end{align}
{ We should note that the absence of the ghost in the model (\ref{FRGBg24}) has been established in \cite{Nojiri:2018ouv, Nojiri:2021mxf}. }

It has been shown that the constraint (\ref{FRGBg20}), which is related to the mimetic condition, is not consistent with the formation of BH horizons \cite{Nojiri:2022cah}.
Therefore, we need to introduce a function $\omega$ in the term of the mimetic constraint so that the resulting field equations can describe the construction
of the BH horizon \cite{Nojiri:2022cah}.
Applying this philosophy, we rewrite the action (\ref{FRGBg22}) in the following form,
\begin{align}
\label{FRGBg222}
S=\int d^4x\sqrt{-g} \left(\frac{1}{2\kappa^2}R
+ \lambda \left( \frac{1}{2}\omega(\chi) \partial_\mu \chi \partial^\mu \chi + \frac{\mu^4}{2} \right)
+ h\left( \chi \right) \mathcal{G} - \tilde V\left( \chi \right)
+ \mathcal{L}_\mathrm{matter}\right)\, .
\end{align}
Variations of the action (\ref{FRGBg222}) w.r.t. the Lagrange multiplier $\lambda$, the auxiliary field $\chi$, and the metric give,
\begin{align}
\label{FRGBg23}
 0=&\frac{1}{2}\omega(\chi) \partial_\mu \chi \partial^\mu \chi + \frac{\mu^4}{2}\,,\\
 \label{FRGBg23gg}
0 =& - \frac{1}{\sqrt{-g}} \partial_\mu \left( \lambda \omega(\chi) g^{\mu\nu}\sqrt{-g}
\partial_\nu \chi \right)
+ h'\left( \chi \right) \mathcal{G} - {\tilde V}'\left( \chi \right) +\frac{1}{2} \lambda \omega'(\chi) g^{\mu\nu}
\partial_\mu \chi\partial_\nu \chi\, ,\\
\label{FRGBg24BB}
0 =& \frac{1}{2\kappa^2}\left(- R_{\mu\nu}
+ \frac{1}{2}g_{\mu\nu} R\right) + \frac{1}{2} T_{\mathrm{matter}\, \mu\nu}
 - \frac{1}{2} \lambda \omega(\chi) \partial_\mu \chi \partial_\nu \chi
 - \frac{1}{2}g_{\mu\nu} \tilde V \left( \chi \right)
+ D_{\mu\nu}^{\ \ \tau\eta} \nabla_\tau \nabla_\eta h \left( \chi
\right)\, .
\end{align}
In the following, we forget the matter energy-momentum tensor because we are interested in vacuum solution.
We are going to apply the field equations (\ref{FRGBg23}), (\ref{FRGBg23gg}), and (\ref{FRGBg24BB}) to a spherically symmetric space-time.

\section{Spherically symmetric BH solutions in Ghost-free $f\left( \mathcal{G} \right)$ Gravity}\label{S3}

In this section, we will study the spherically symmetric space-time created by solving Eqs.~(\ref{FRGBg23}), (\ref{FRGBg23gg}) and (\ref{FRGBg24BB})
given by the ghost-free $f\left( \mathcal{G} \right)$ gravitational theory defined by (\ref{FRGBg222}).
Specifically, we investigate if it is possible to derive spherically symmetric BH solutions.

\subsection{Schwarzshild-type black hole solutions}

Now, we investigate how the field equations for the theory (\ref{FRGBg222}) behave in the case of the spherically symmetric metric with the following line-element,
\begin{align}
\label{metric1}
ds^2 = - f(r) dt^2 + \frac{dr^2}{f(r)}{+}r^2 d\Omega^2\,, \quad \mbox{where} \quad d\Omega^2=d\theta^2+\sin^2\theta d\phi^2\,.
\end{align}
For this metric, we have,
\begin{align}
\label{E2}
& \Gamma^r_{tt}=f^2\Gamma^t_{tr}=-f^2\Gamma^r_{rr}=\frac{1}{2}ff'\, ,\quad \Gamma^\theta_{r\theta}=\Gamma^\phi_{r\phi}=\frac{1}{r}\, ,
\quad \Gamma^r_{\theta \theta}=\frac{\Gamma^r_{\phi \phi}}{\sin^2\theta}=-f\,r\, ,\quad
\Gamma^\phi_{\theta \phi}=-\frac{\Gamma^\theta_{\phi \phi}}{\sin^2\theta}=\frac{\cos\theta}{\sin\theta}\,,\nonumber\\
& \mathcal{G}=\frac{4(f'^2+ff''-f'')}{r^2}\,,
\end{align}
where $f'\equiv f'(r) \equiv \frac{df(r)}{dr}$.
Moreover, we assume that $\lambda$, $\omega$, and $\chi$ only depend on the radial coordinate $r$, i.e.,
$\lambda=\lambda(r)$, $\omega=\omega(r)$ and $\chi=\chi(r)$.

Actually, the $(t,t)$-component, $(r,r)$-component, and $(\theta,\theta)=(\phi,\phi)$-components of the field
(\ref{FRGBg24BB}) give,
\begin{align}
\label{W1}
0 = &\, \frac{4h'f'-8f^2h''+r^2V+rf'+8fh''-1-12ff'h'+f}{r^2}\, , \\
\label{W2}
0 = &\, \frac{f+4f'h'+r^2 f\lambda\omega \chi'^2+r^2V+rf'-1-12ff'h'}{r^2}\, ,
\\
\label{W3}
0 = &\, \frac{2rV+2f'+rf''-8h'f'^2-8h'ff''-8ff'h''}{2r}\, .
\end{align}
On the other hand, Eqs.~(\ref{FRGBg23}) and (\ref{FRGBg23gg}) yield,
\begin{align}
\label{W4}
0 =&\, \mu^4+\omega f \chi'^2 \, , \\
\label{W5}
0 =&\, \frac{8h'f''(f-1)-2r^2f\lambda \omega\chi'\chi''-r\chi'^2[2r\lambda \omega\,f'+f\{2r\omega \lambda'+\lambda(4\omega+r\omega')\}]+8f'^2h'-2rV' }{2r^2\chi'}\,.
\end{align}
Equations~(\ref{W1})-(\ref{W5}) are five non-linear differential equations in six unknown functions $f$, $h$, $V$, $\lambda$, $\omega$, and $\chi$, therefore,
we are going to fix some of these unknown functions to derive the other ones.
First, we solve Eq.~(\ref{W4}) and obtain
\begin{align}
\label{W6}
%f=1+\frac{\alpha r^2}{\beta +r^3}
\chi=c_0 r \quad \Rightarrow \quad \omega=-\frac{\mu^4}{c_0{}^2f}\, .
\end{align}
Substituting Eq.~(\ref{W6}) into Eqs.~(\ref{W1})-(\ref{W5}), we obtain
\begin{align}
\label{W7}
f=&\, 1-\frac{2M}{r}+\frac{c_1}{r^2}+\frac{c_2}{r^6}\,,\nonumber\\
V =&\, \frac{1}{\left( 3Mr^{20}-2r^{19}c_1-4r^{15} c_2 \right)}\Biggl\{ 2\Upsilon(r)\Upsilon_1(r)
\int \frac{r^6 \left( c_1r^4+10c_2 \right)}{ \Upsilon_1(r)\left(2Mr^5 -r^6-c_1r^4-c_2 \right) \left( 3Mr^5-2c_1r^4-4c_2 \right)}{dr} \nonumber\\
&\, \qquad -8c_3 \Upsilon(r) \Upsilon_1(r) - r^{11} \left( r^5Mc_1+25rMc_2-8c_1c_2 \right) \Biggr\}
\, ,\nonumber \\
h=&\, c_4+\int \Upsilon_1(r) \left( \int \frac{r^6 \left( c_1r^4+10c_2 \right)}{ 4 \Upsilon_1(r)\left(r^6+c_1r^4+c_2-2\,M r^5 \right)
\left( 3Mr^5-2 c_1 r^4-4c_2 \right)}{dr}-4\,c_3 \right) { dr}\,,\nonumber\\
\lambda=&\, \frac{2\left(2Mr^5 -c_2-c_1 r^4 \right)}{r^{15}\mu^4 \left( 3 Mr^5-2 c_1 r^4-4\,c_2 \right)} \Biggl\{4 \Upsilon_1(r) \Upsilon_2(r)
\int \frac{r^6\left( c_1 r^4+10 c_2 \right)}{ \Upsilon_1(r) \left(2 Mr^5 -r^6-c_1 r^4-4c_2 \right) \left( 3 Mr^5-2 c_1 r^4-4 c_2 \right)}{dr}\nonumber\\
&\, \qquad -c_3 \Upsilon_1(r)\Upsilon_2(r)-\left( c_1 r^4+10 c_2 \right) r^7 \Biggr\} \,,
\end{align}
where
\begin{align}
\Upsilon(r)=&\, \left(2 M^2r^{16}-2M \left( 2c_1+3 M^2 \right) r^{15}+ \left( 15 M^2c_1+{c_1}^2 \right) r^{14}-10 Mr^{13}{c_1}^2+2{c_1}^3r^{12}
+ \left( 79 M^2c_2 +12 c_1 c_2 \right) r^{10} \right. \nonumber\\
&\, \left. -32Mr^{11}c_2-80Mr^9c_1 c_2+20{c_1}^2r^8c_2+3{c_2}^2 r^6-62Mr^5{c_2}^2+30c_1 r^4{c_2}^2+12 {c_2}^3 \right)\,,\nonumber\\
\Upsilon_1(r)=&\, \e^{\int \frac{\left( 12 M^2+5 c_1 \right) r^{10}-4 M r^{11}-21 Mr^9c_1+8 {c_1}^2r^8+27 c_2 r^6-77 Mr^5c_2+48 c_1 r^4c_2
+48 {c_2}^2}{r \left(2 Mr^5 - r^6-c_1 r^4-c_2 \right) \left( 3 Mr^{ 5}-2 c_1 r^4-4 c_2 \right) }dr} \,,\nonumber\\
\Upsilon_2(r)=&\, 4Mr^{11}- \left(12 M^2+5 c_1\right) r^{10}+21Mr^9c_1-8{c_1}^2r^8-27c_2 r^6+77 Mr^5c_2-48 c_1 r^4c_2-48 {c_2}^2 \, .
\end{align}
The curvature invariants associated with solution (\ref{W6}) take the following form,
\begin{align}
\label{W8}
K=&\,R_{\alpha\beta\gamma\rho}R^{\alpha\beta\gamma\rho}=\frac{48M^2}{r^6}-\frac{96Mc_1}{r^7}
+\frac{56{c_1}^2}{r^8}-\frac{488Mc_2}{r^{11}}+\frac{608c_1c_2}{r^{12}}+\frac{1912{c_2}^2}{r^{16}}\,, \nonumber \\
R_{\alpha\beta}R^{\alpha\beta}=&\, \frac{4{c_1}^2}{r^8}+\frac{80c_1c_2}{r^{12}}+\frac{500{c_2}^2}{r^{16}}\,, \quad
R=- \frac{20}{r^8}\,, \\
\label{G11}
\mathcal{G}=&\, \frac{8 \left( 6M^2r^{10}-12Mc_1r^9-56Mc_2r^5+5{c_1}^2r^8+36c_1c_2r^4+39{c_2}^2 \right)}{r^{16}}\,.
\end{align}
Eq.~(\ref{W8}) shows that the BH solution given by Eq.~(\ref{W6}) has a { hard} singularity when $r\to 0$
compared with the Schwarzschild solution of GR \cite{Nashed:2021mpz} where the Kreschmann scalar $K$ behaves as $K\sim r^{-6}$.

\subsection{More general black hole}

Now, let us investigate how the field equations in the
theory (\ref{FRGBg222}) behave in the case of a spherically symmetric metric with the following line element:
\begin{align}
\label{metric2}
ds^2 = - f(r) dt^2 + \frac{dr^2}{f_1(r)}-r^2 d\Omega^2 \quad \mbox{where} \quad d\Omega^2=d\theta^2+\sin^2\theta d\phi^2\,.
\end{align}
For this metric, we have,
\begin{align}
\label{E2B}
\Gamma^r_{tt}=&\, ff_1\Gamma^t_{tr}=-{f_1}^2\Gamma^r_{rr}=\frac{1}{2}f_1f'\, ,\quad \Gamma^\theta_{r\theta}=\Gamma^\phi_{r\phi}=\frac{1}{r}\, ,
\quad \Gamma^r_{\theta \theta}=\frac{\Gamma^r_{\phi \phi}}{\sin^2\theta}=-f_1\,r\, ,\quad
\Gamma^\phi_{\theta \phi}=-\frac{\Gamma^\theta_{\phi \phi}}{\sin^2\theta}=\frac{\cos\theta}{\sin\theta}\,,\nonumber\\\mathcal{G}=&\, \frac{2(f_1[1-f_1]f'^2+f[3f_1-1]f'f'_1-2ff_1f''[1-f_1])}{r^2f^2}\,,
\end{align}
where $f'_1\equiv f'_1(r) \equiv \frac{df_1(r)}{dr}$.
We assume that $\lambda$, $\omega$, and $\chi$ only depend on the radial coordinate $r$, i.e.,
$\lambda=\lambda(r)$, $\omega=\omega(r)$ and $\chi=\chi(r)$, again.

The $(t,t)$-component, $(r,r)$-component, and $(\theta,\theta)=(\phi,\phi)$-components of the field equation
(\ref{FRGBg24BB}) have the following forms,
\begin{align}
\label{W11}
0 = &\, \frac{f_1-1+rf'_1+8f_1h''-8{f_1}^2h''+4f'_1h'+r^2V-12f_1f'_1h'}{r^2}\, , \\
\label{W22}
0 = &\, \frac{ff_1+4f_1f'h'+r^2 ff_1\lambda\omega \chi'^2+r^2fV+rf_1f'-f-12{f_1}^2f'h'}{fr^2}\, , \\
\label{W33}
0 = &\, \frac{2f^2f'_1+4rf^2V+2ff_1f'-24ff_1h'f'f'_1-rf_1f'^2+2rff_1f''+rff'f'_1+8{f_1}^2f'^2h'-16{f_1}^2h'f''-16f{f_1}^2f'h''}{4rf^2}\, .
\end{align}
On the other hand, Eqs.~(\ref{FRGBg23}) and (\ref{FRGBg23gg}) yield,
\begin{align}
\label{W44}
0 =&\, \mu^4+\omega f_1 \chi'^2 \, ,  \\
\label{W55}
0 =&\, \frac{1}{2r^2f^2\chi'}\left\{8ff_1h'f'' \left( f_1-1 \right) -2r^2f^2f_1\lambda \omega\chi'\chi''
 -rf\chi'^2 \left[f_1r\lambda \omega f'+f \left\{r\lambda \omega f'_1+f_1 \left[ 2r\omega \lambda'+\lambda \left( 4\omega+r\omega' \right) \right] \right\} \right] \right. \nonumber\\
&\, \left. \qquad +4f'h'\left[ f'f_1 \left( 1-f_1 \right) +ff'_1 \left( 4f_1-1 \right) \right] -2r^2f^2V' \right\}\,.
\end{align}
Equations~(\ref{W11})-(\ref{W55}) are five non-linear differential equations in seven unknown functions $f$, $f_1$, $h$, $V$, $\lambda$, $\omega$, and $\chi$,
therefore, we are going to fix some of these unknown functions to derive the other ones.
By using Eq.~(\ref{W44}), we obtain
\begin{align}
\label{W66}
\chi=c_0 r \quad \Rightarrow \quad \omega=-\frac{\mu^4}{{c_0}^2f_1}\, .
\end{align}
By substituting Eq.~(\ref{W66}) into Eqs.~(\ref{W11})-(\ref{W55}), we obtain
\begin{align}
\label{W77}
f=&\, 1+\frac{\alpha r^2}{\beta +r^3}\,, \quad
f_1=1+\frac{\alpha}{r}+\frac{\beta}{r^3}\,,\nonumber\\
V =&\, \frac{1}{4r^7 \left( r^3+\beta \right) \left( 3 \alpha r^5+2 r^3\beta+2 \beta^2 \right)} \Biggl\{ \beta \alpha \Upsilon_3(r)\Upsilon_4(r)
\int \frac{r^3 \left(2 \beta^{ 2} -27 \alpha r^5+4 r^3\beta+2 r^6 \right)}{\Upsilon_3(r)\left( r^3+\beta \right) \left( 3\alpha r^5+2 r^3\beta+2 \beta^2 \right)
\left( r^3+\beta+\alpha r^2 \right)}{dr} \nonumber\\
&\, -16 c_5  \Upsilon_4(r) \alpha \Upsilon_3(r) -4 r^4\beta^3\alpha+70 r^7\beta^2\alpha+20 r^{10} \beta \alpha
+12 r^2\beta^4+24 r^5\beta^3+12 r^8\beta^2+54 \beta r^9\alpha^2 \Biggr\}\, ,\nonumber\\
h=&\, c_6+\int \left( \beta \int \frac{r^3 \left( 27 \alpha r^5-2 \beta^2-4 r^3\beta-2 r^6 \right)}
{16\Upsilon_3(r) \left( r^3+\beta \right) \left( 3 \alpha r^5+2 r^3\beta+2 \beta^2 \right)
\left( r^3+\beta+\alpha r^2 \right)}{dr}-c_5 \right) \Upsilon_3(r){dr}\,,\nonumber\\
\lambda=&\, \frac{1}{2\mu^4 \left( r^3+ \beta \right)r^7 \left(3 \alpha r^5+2r^3\beta+2\beta^2 \right)} \Biggl\{ \beta \alpha \Upsilon_3(r) \Upsilon_5(r)
\int \frac{r^3 \left(2 \beta^2 -27 \alpha r^5+4 r^3\beta+2 r^6 \right)}{\Upsilon_3(r) \left( r^3+\beta \right)\left( 3\,\alpha r^5+2 r^{ 3}\beta+2 \beta^2 \right)
\left( r^3+\beta+\alpha r^2 \right) }{dr}\nonumber\\
&\, -16\,\alpha c_5 \Upsilon_3(r) \Upsilon_5(r)+16 r^{10} \beta \alpha+10 r^8\beta^2+45 \beta r^9\alpha^2+10 r^2\beta^4
+20 r^5\beta^3+10 r^4\beta^3 \alpha+53 r^7\beta^2\alpha \Biggr\} \,,
\end{align}
where
\begin{align}
\Upsilon_3(r)=&\, \e^{\int \frac{12 \alpha^2r^{10}-15 \alpha^2 r^7\beta+10 \beta^3\alpha r^2+ 19 r^5\alpha \beta^2+17 r^8\alpha \beta
+8 r^{11} \alpha+18 \beta^4+48 r^3\beta^3+42 r^6\beta^2+ 12 r^9\beta}{2 \left( 3 \alpha r^5+2 r^3\beta+2 \beta^2 \right)
\left( r^3+\beta+\alpha r^2 \right) r \left(  r^3+\beta \right) }dr} \,,\\
\Upsilon_4(r)=&\, 12 r^3\beta^3-2r^{11}\alpha-3\alpha^2r^{10}-4\beta^3\alpha r^2+44r^5\alpha\beta^2-4\beta^4+18r^6\beta^2+19r^8\alpha \beta+24\alpha^2r^7\beta+2r^9\beta \,,\\
\Upsilon_5(r)=&\, 8r^6\beta^2+8r^8\alpha \beta+21\alpha^2r^7\beta-r^9\beta+13r^3\beta^3+4\beta^3\alpha r^2+34r^5\alpha \beta^2+4\beta^4-4r^{11}\alpha-6 \alpha^2r^{10}\,,
\end{align}
and the functions $\chi$ and $\omega$ have the same form as given by Eq.~(\ref{W7}).
Calculating the curvature invariants of solution (\ref{W77}) we obtain
\begin{align}
\label{W88}
K=&\, R_{\alpha\beta\gamma\rho}R^{\alpha\beta\gamma\rho}
=\frac{3}{4r^{10}\left( r^3+\beta \right)^4}\left\{16\alpha^2r^{16}-40\alpha^2\beta r^{13}+243\alpha^2\beta^2r^{10}+76\alpha^2\beta^3r^7
+20\alpha^2\beta^4r^4+80\alpha\beta r^5+44\beta^6 \right. \nonumber\\
&\, \left. +66\beta^2r^{12}+66\beta^3r^9+156\beta^4r^6+80\alpha \beta r^{14}+80\alpha\beta^2r^{11}+160\alpha\beta^3r^8+196\beta^5r^3+80\alpha\beta^5r^2\right\}
%\approx\frac{12\alpha^2}{r^6}-\frac{120\alpha^2\beta}{r^9}+\frac{756\alpha^2\beta^2}{r^{12}}
\,, \\
R_{\alpha\beta}R^{\alpha\beta}=&\, \frac{\beta^2}{8r^{10} \left( r^3+\beta \right)^4}\left\{116\beta^2r^6+40\beta^3 r^3+19\beta^4+801\alpha^2r^{10}
+76r^{12}+93\alpha r^{11}+76\beta r^9+198\beta\alpha r^8 \right. \nonumber\\
&\, \left. +117\alpha \beta^3r^2+112\alpha^2 \beta r^7+88\alpha^2\beta^2r^4\right\}
%\approx\frac{90\alpha^2\beta^2}{r^{12}}-\frac{576\alpha^2\beta^3}{r^{15}}+\frac{2142\alpha^2\beta^4}{r^{18}}
\,, \\
R=&\, \frac{\beta \left( 8r^6+21\alpha r^5+16\beta r^3-6\alpha \beta r^2+8\beta^2 \right)}{2r^5 \left(r^3+\beta\right)^2}
%\approx\frac{6\alpha\beta}{r^6}-\frac{30\alpha\beta^2}{r^9}+\frac{72\alpha\beta^3}{r^{12}}
\,, \\
\label{G33}
\mathcal{G}=&\, \frac{2\alpha \left( 6\alpha r^8+10\beta r^6-27\beta\alpha r^5-31\beta^2 r^3-14\beta^3-6\alpha\beta^2 r^2 \right)}
{2r^5 \left( r^3+\beta \right)^2}
%\approx\frac{12\alpha^2}{r^6}+\frac{20\alpha\beta}{r^8}-\frac{78\alpha^2\beta}{r^9}
\,.
\end{align}
{ The above invariants show that there is no singularity at $r=0$. }
%also a soft singularity of the BH solution (\ref{W66}) compared with the Schwarzschild solution of GR.

\section{Relevant physics and thermodynamics of the BHs (\ref{W6}, \ref{W7}) and (\ref{W66}, \ref{W77})}\label{S4}

In this section, we are going to investigate the essential physics of solutions (\ref{W6}, \ref{W7}) and (\ref{W66}, \ref{W77}).
\subsection{Relevant physics and thermodynamics of the BH (\ref{W6}, \ref{W7})}
\begin{figure}
\centering
\subfigure[~The plot of the function $f(r)$, given by Eq. \eqref{W7}, vs. the radial coordinate $r$ ]{\label{fig:1a}\includegraphics[scale=0.25]{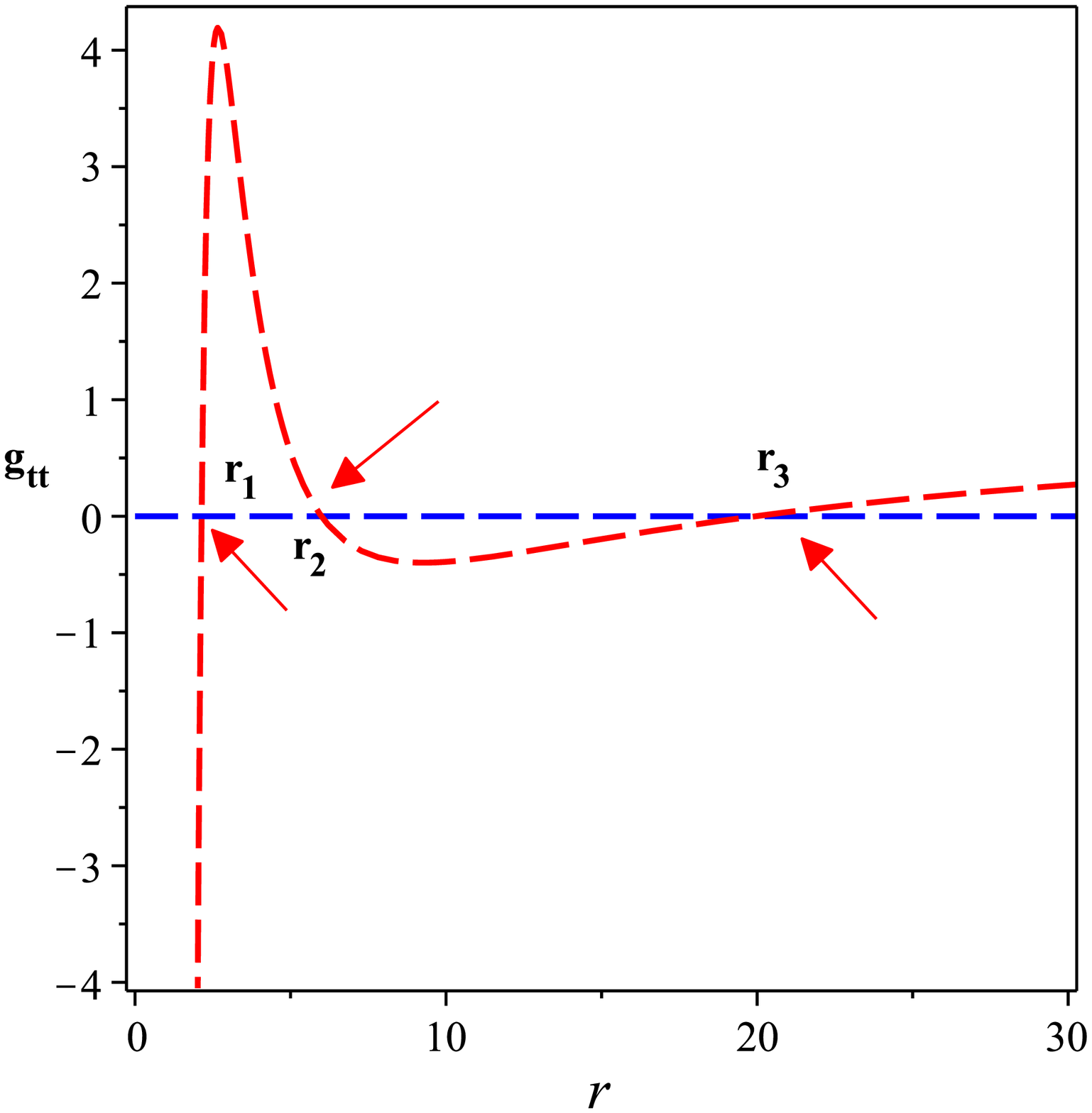}}\hspace{0.2cm}
\subfigure[~The plot of GB invariant given by Eq.~(\ref{W8}) vs. the radial coordinate $r$]{\label{fig:1b}\includegraphics[scale=0.25]{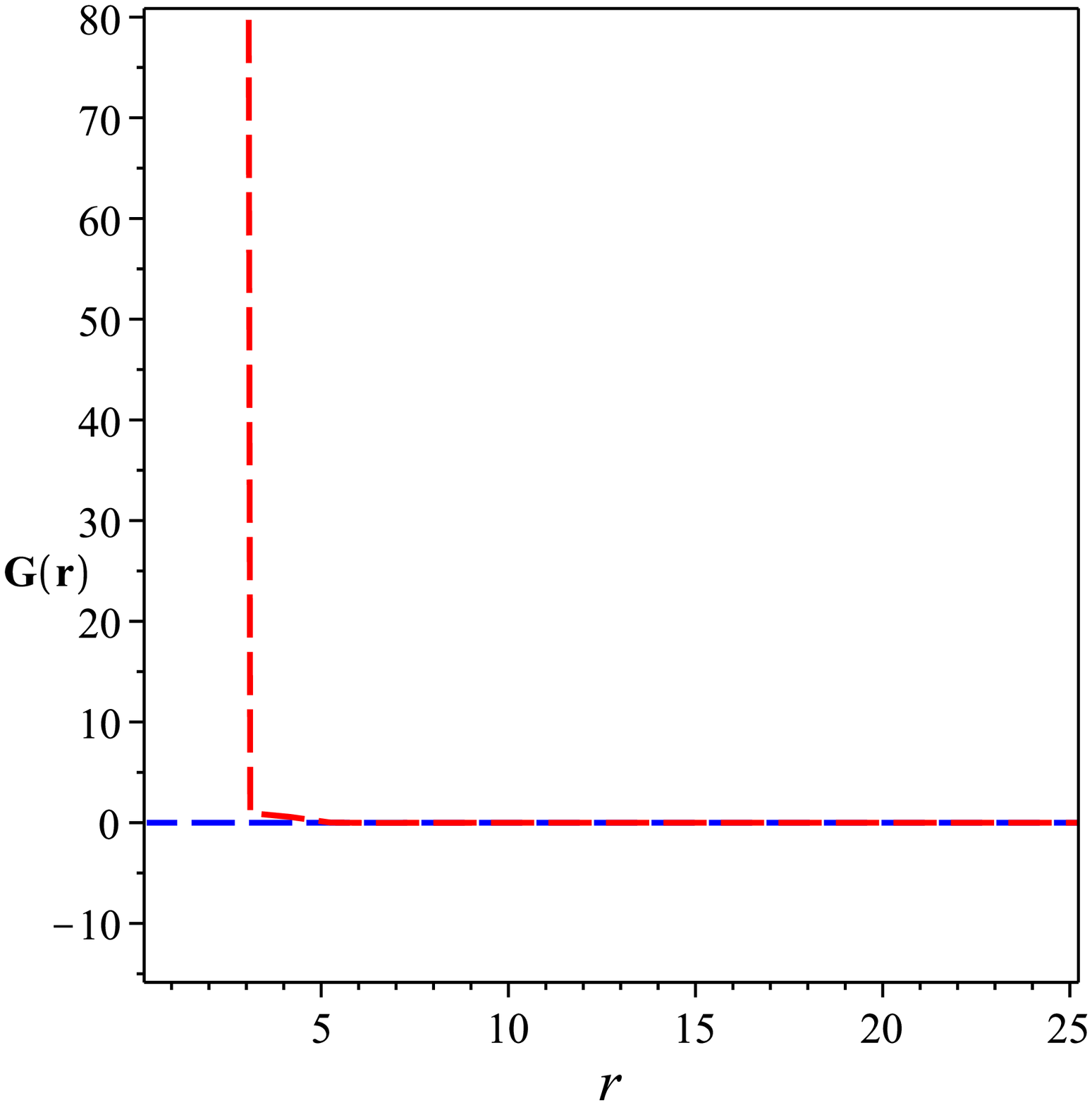}}\hspace{0.2cm}
\subfigure[~The function $h(\chi)$ vs. $\chi$ ]{\label{fig:1c}\includegraphics[scale=0.25]{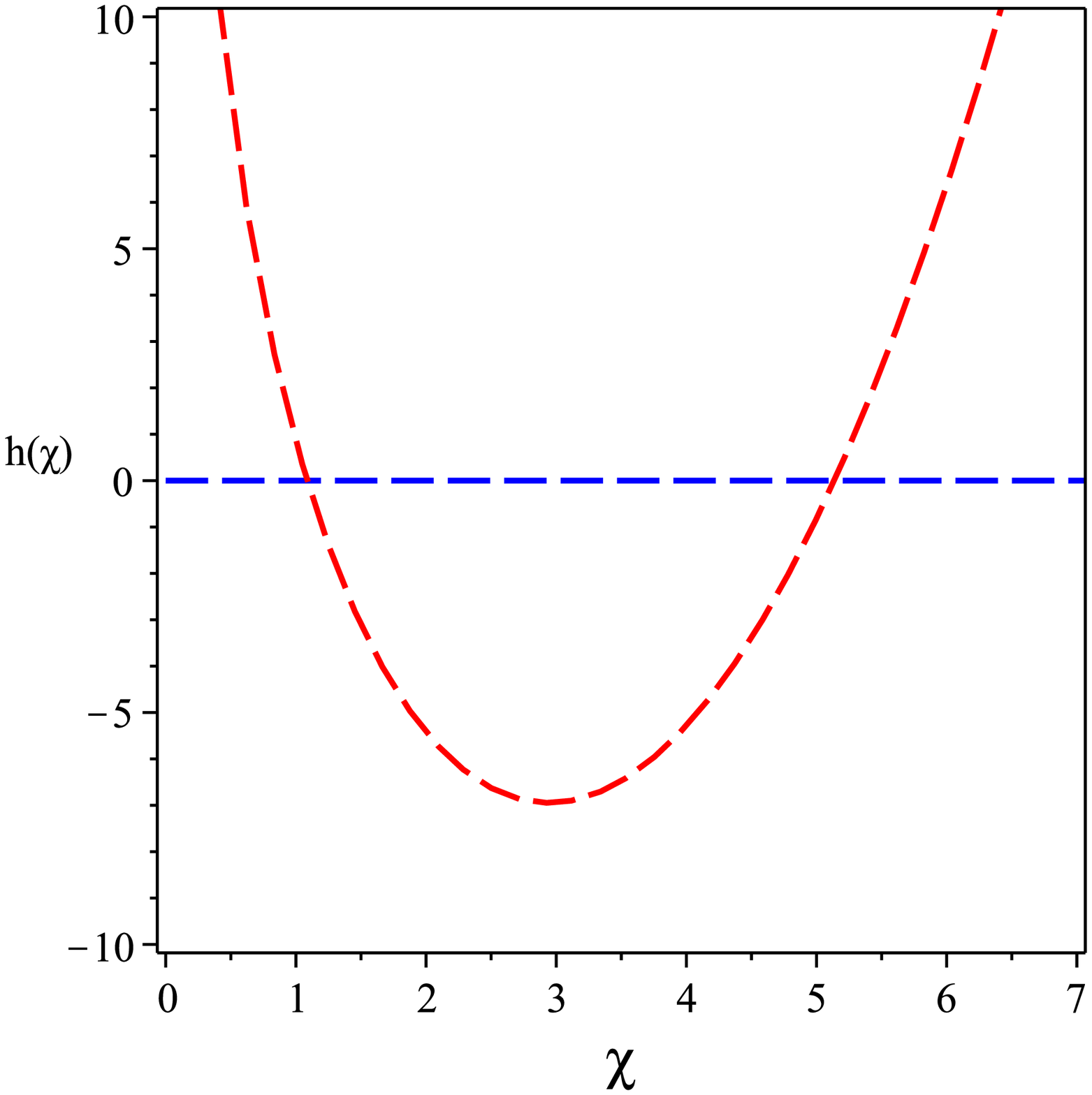}}\hspace{0.2cm}
\subfigure[~The potential $V(\chi)$ vs. $\chi$ ]{\label{fig:1d}\includegraphics[scale=0.25]{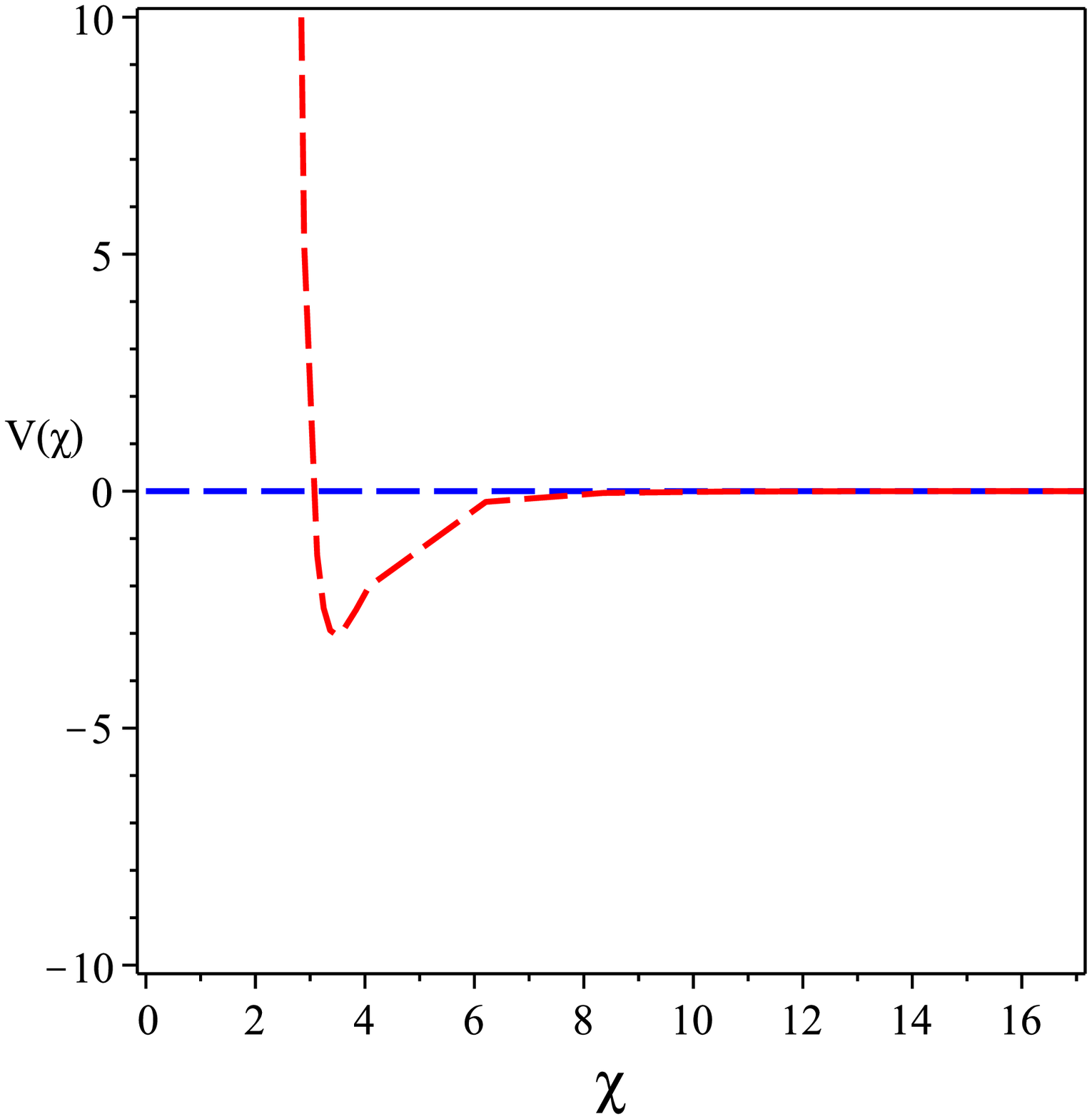}}
\subfigure[~The Lagrangian potential $\lambda$ vs. $r$ ]{\label{fig:1e}\includegraphics[scale=0.25]{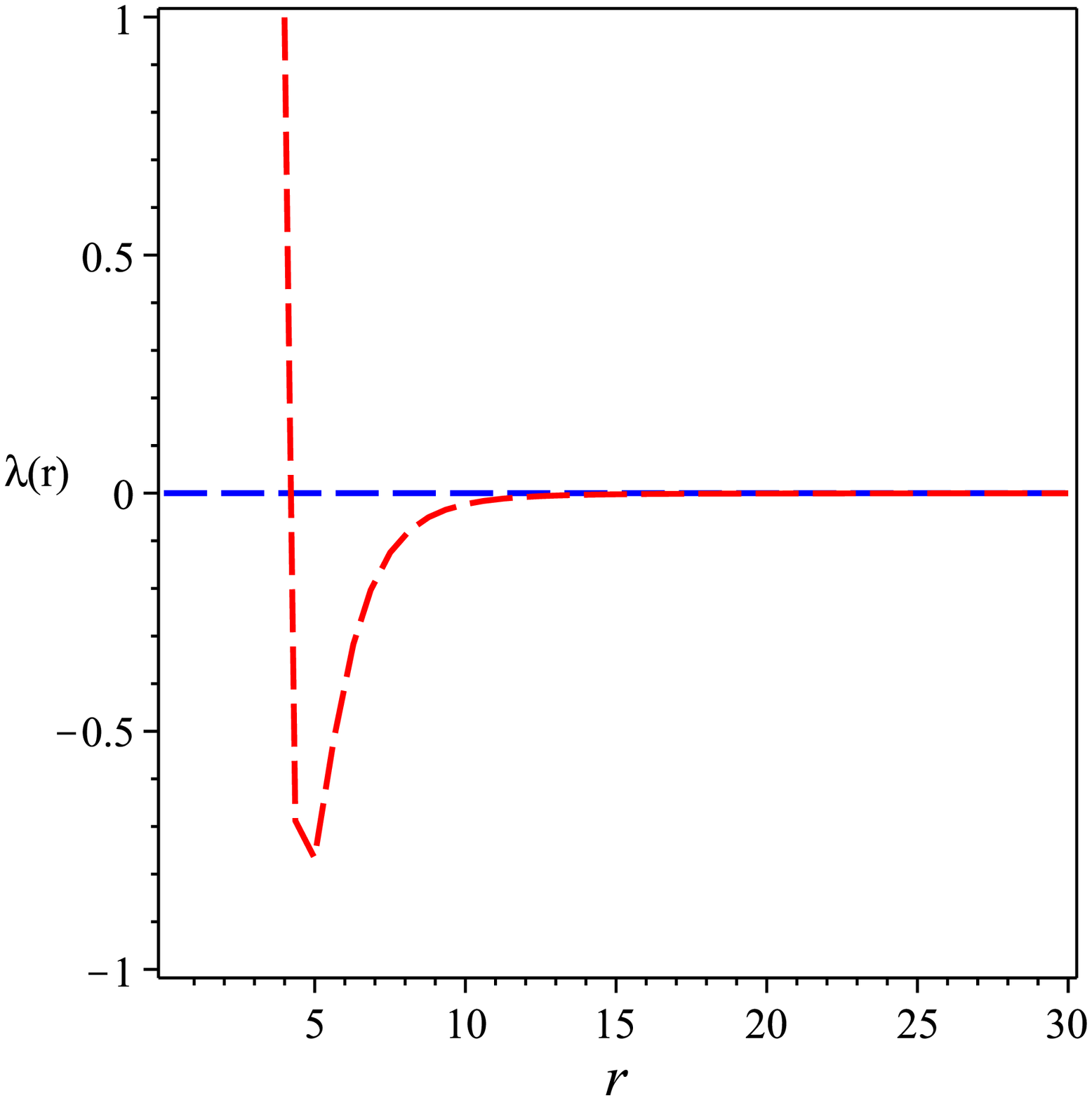}}
\caption{Schematic plots of the radial coordinate $r$  \subref{fig:1a} vs. the function $f$ given by Eq. \eqref{W7};  \subref{fig:1b} vs. the GB invariant given by Eq.~(\ref{G11}); \subref{fig:1c} the function $h$ vs. $\chi$; \subref{fig:1d} the potential $V$ vs. $\chi$, and \subref{fig:1e} the Lagrange multiplier $\lambda(r)$ given by Eq. (\eqref{W7}) vs. $r$.}
\label{Fig:1}
\end{figure}

For the the BH (\ref{W7}), we are going to write the line-element as,
\begin{align}
\label{metaf1}
ds^2=-\left[1-\frac{2M}{r}+\frac{c_1}{r^2}+\frac{c_2}{r^6}\right]dt^2+\frac{dr^2}{ 1-\frac{2M}{r}+\frac{c_1}{r^2}+\frac{c_2}{r^6}}
+r^2 \left( d\theta^2+\sin^2\theta d\phi^2 \right)\,.
\end{align}
{ The metric of the line-element (\ref{metaf1}) has multi-horizons as FIG.~\ref{Fig:1}~\subref{fig:1a} shows. These multi-horizons, three ones, are created due to specific value of the constant $c_2$ and other values will create two horizons only.}
These three multi-horizons are created from the constants $M$, $c_1$, and $c_2$ and the vanishing of the dimensional parameter $c_2$ reproduces geometry with two horizons.
Moreover, when the dimensional parameters $c_1$ and $c_2$ vanish, the geometry with one horizon, i.e., the Schwarzschild geometry is reproduced.
As Eq.~(\ref{G11}) shows, the BH solution (\ref{metaf1}) gives a non-trivial form of the GB invariant, whose behavior is shown in FIG.~\ref{Fig:1}~\subref{fig:1b}.
The behavior of the physical quantities, $h(\chi)$, $V(\chi)$, and the Lagrange multiplier field $\lambda$, for the BH solution (\ref{W7}) are shown
in FIG.~\ref{Fig:1}~\subref{fig:1c}, \ref{Fig:1}~\subref{fig:1d}, and \ref{Fig:1}~\subref{fig:1e}.
Using Eq.~(\ref{metaf1}), we obtain $M$ as a function of the redial coordinate $r$,
\begin{align}
\label{m2}
M=\frac{r}{2}\left(1+\frac{c_1}{r^2}+\frac{c_2}{r^6}\right)\,.
\end{align}

Now we investigate the thermodynamics for the BH (\ref{W7}).
The Hawking temperature is defined as \cite{Sheykhi:2012zz,Sheykhi:2010zz,Hendi:2010gq,Sheykhi:2009pf}
\begin{align}
\label{temp}
T_2 = \frac{f'\left( r_2 \right)}{4\pi}\,,
\end{align}
where $r_2$ is the event horizon located at $r = r_2$ which is the largest positive root of $f\left( r_2 \right) = 0$ which satisfies $f'\left( r_2 \right)\neq 0$.
Using Eq.~(\ref{temp}), we obtain the Hawking temperature of the BH solution in the form:
\begin{align}
\label{temp11}
T_2 =\frac{M {r_2}^5-c_1 {r_2}^4-3c_2}{2\pi {r_2}^7}\,.
\end{align}
The Hawking entropy is defined as \cite{Cognola:2011nj,Sheykhi:2012zz,Sheykhi:2010zz,Hendi:2010gq,Sheykhi:2009pf,Zheng:2018fyn}
\begin{align}\label{ent}
S\left( r_2 \right)=\frac{1}{4}A\left( r_2 \right)\, ,
\end{align}
where $A$ is the area of the event horizon.

\begin{figure}
\centering
\subfigure[~{~{The horizons, $r_1$ and $r_1$,  of the BH solution (\ref{W6})}}]{\label{fig:2a}\includegraphics[scale=0.25]{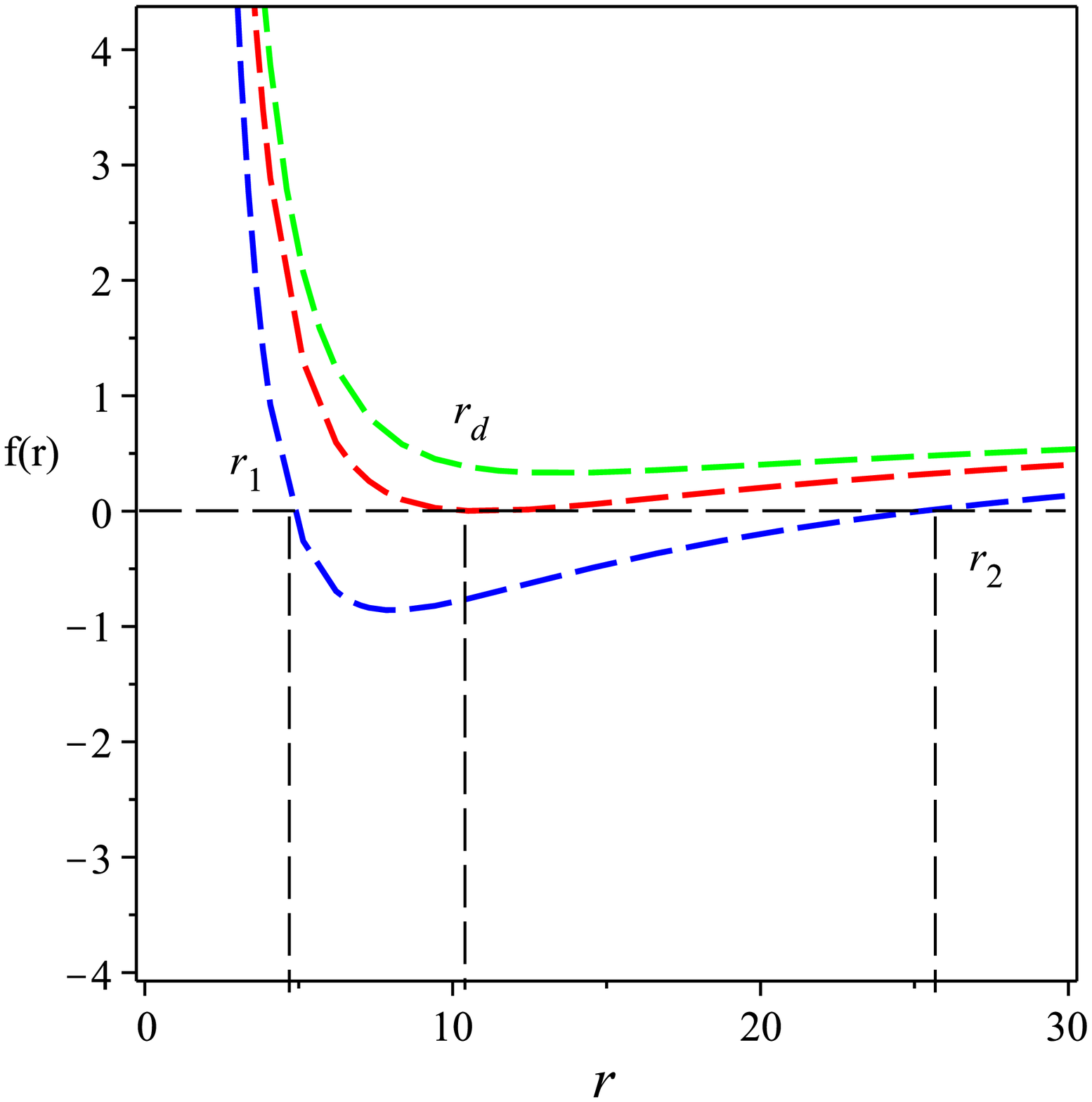}}\hspace{0.2cm}
\subfigure[~The entropy of the black hole solution (\ref{W6})~{ vs. $r_2$} ]{\label{fig:2b}\includegraphics[scale=0.25]{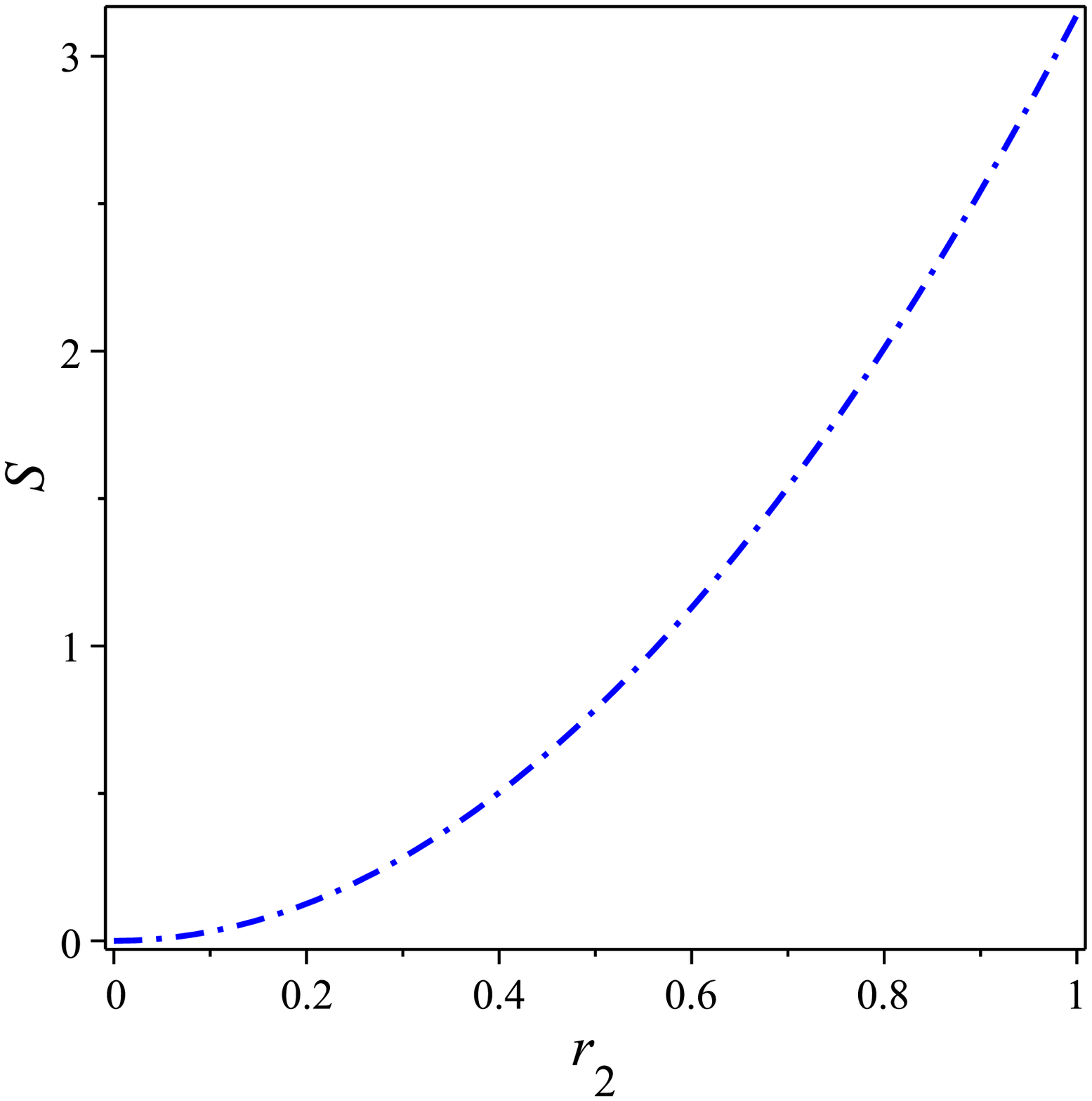}}\hspace{0.2cm}
\subfigure[~The temperature of the BH solution (\ref{W6})~{ vs. $r_2$ } ]{\label{fig:2c}\includegraphics[scale=0.25]{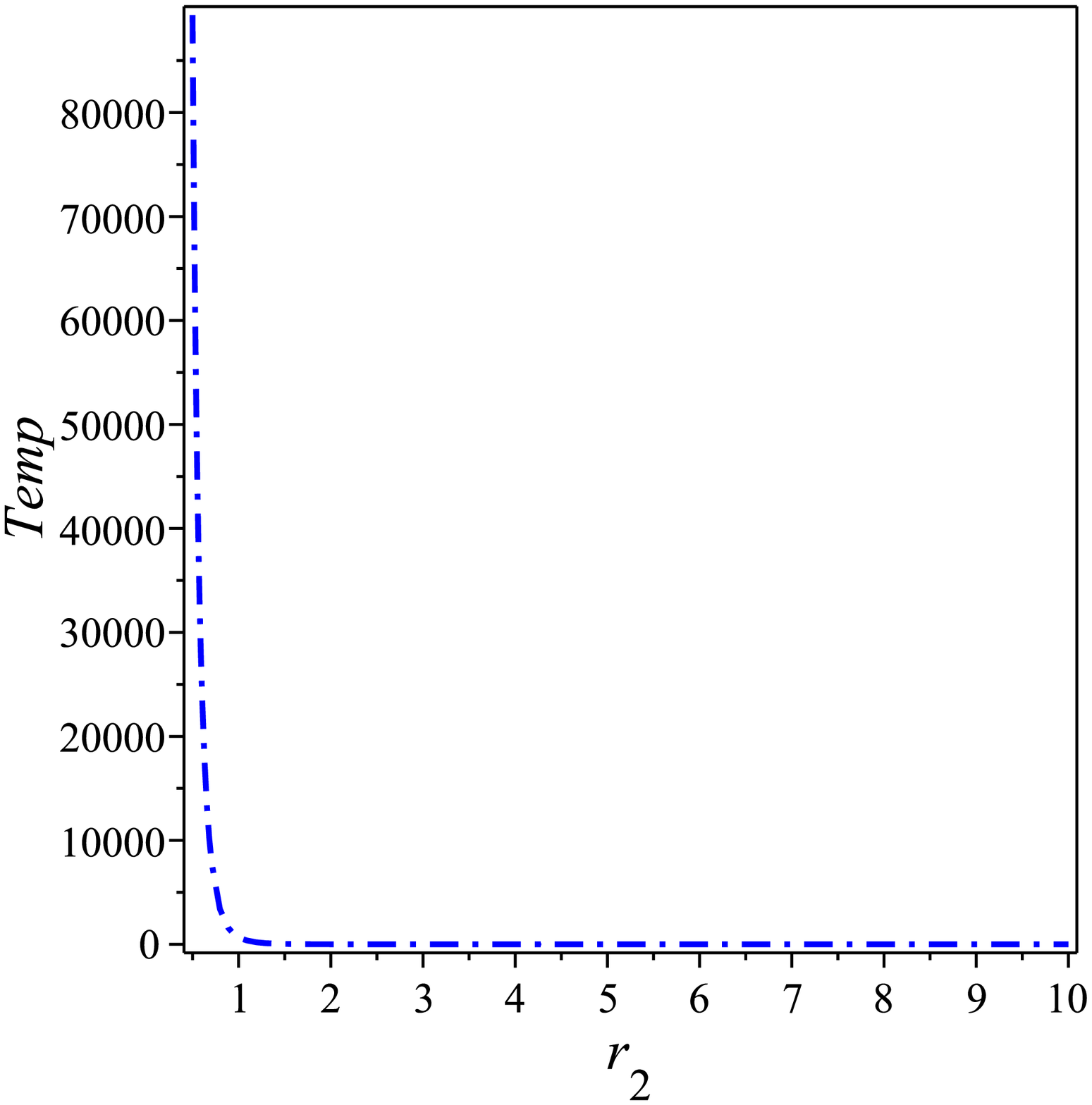}}\hspace{0.2cm}
\subfigure[~The heat capacity of the BH solution (\ref{W6}) vs.~{ $r_2$}]{\label{fig:2d}\includegraphics[scale=0.25]{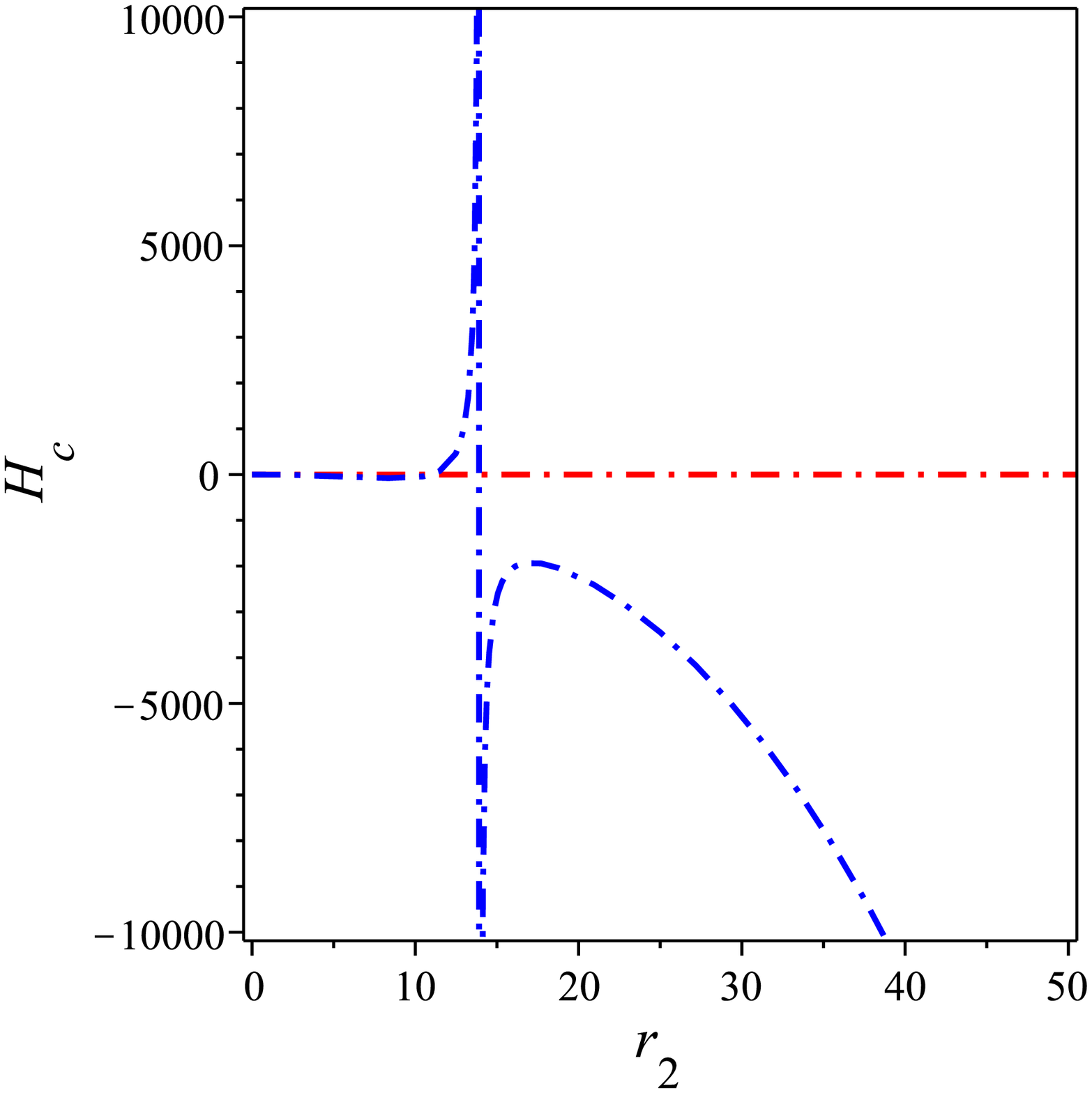}}
\subfigure[~{ The Gibbs free energy of the BH solution (\ref{W6}) vs. $r_2$}]{\label{fig:2e}\includegraphics[scale=0.25]{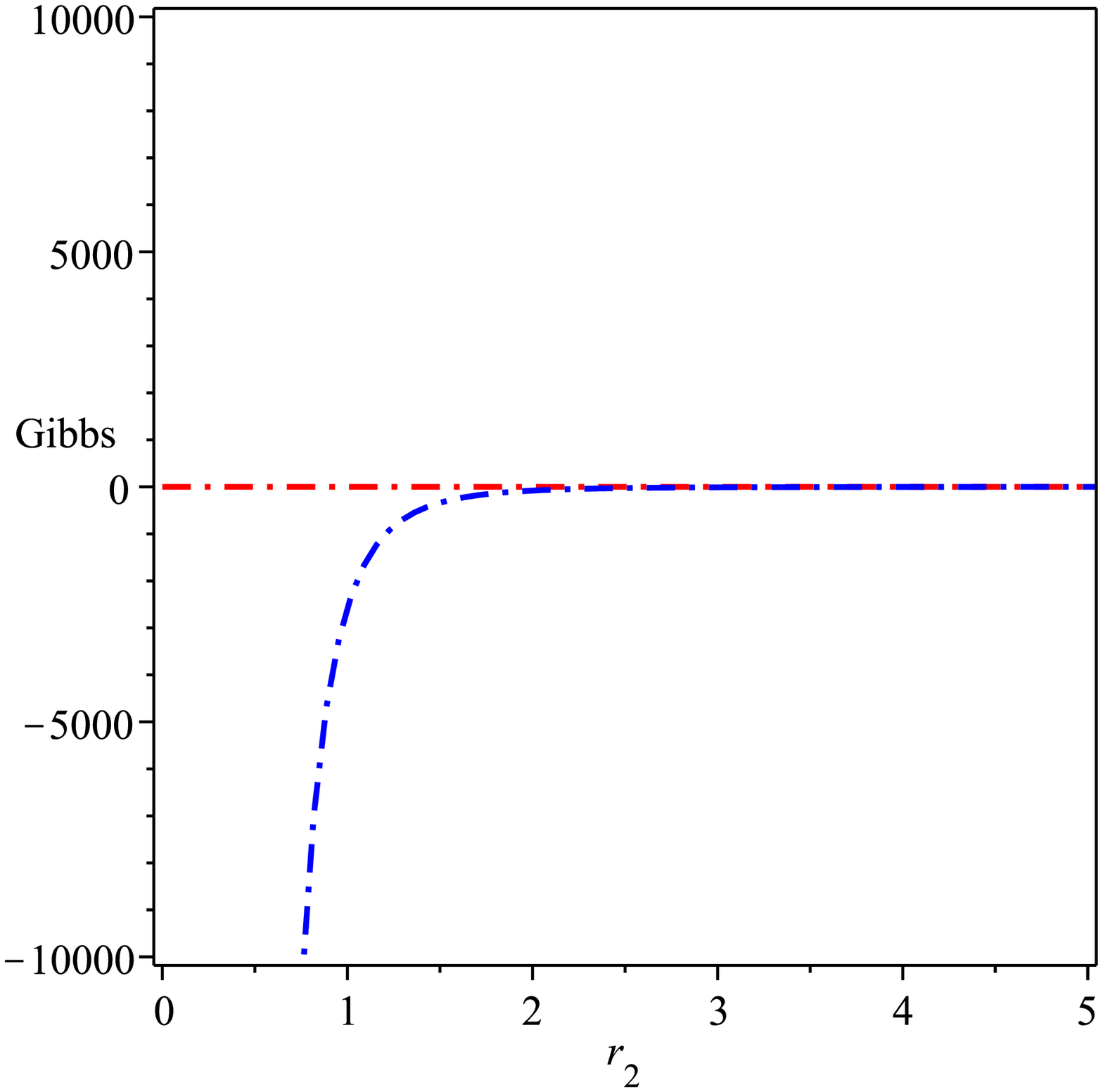}}
\caption{{ { Schematic plot of \subref{fig:2a} the horizons, $r_1$ and $r_2$,  of the BH solution (\ref{W6})}; \subref{fig:2b} the entropy of the BH solution (\ref{W6}); \subref{fig:2c} the Hawking temperature of the BH solution (\ref{W6}); \subref{fig:2d} the heat capacity of the BH solution (\ref{W6}) finally, \subref{fig:2e} the Gibbs free energy of the BH solution (\ref{W6}).}}
\label{Fig:2}
\end{figure}

{ To show how many horizons of the BH solution Eq.~(\ref{W6}), we plot $g_{00}$ in FIG.~\ref{Fig:2}~\subref{fig:2a}.
As FIG.~\ref{Fig:1}~\subref{fig:1a} shows that for specific value of  $c_2$, we have three horizons and for other values  $c_2$ or when $c_2=0$   we have two horizons as FIG.~\ref{Fig:2}~\subref{fig:2a}.
Also in FIG.~\ref{Fig:2}~\subref{fig:2a}, we show the region where the black hole has no singularity, i.e., naked singularity.}

Using Eq.~(\ref{ent}), the entropy of the BH (\ref{W6}) is computed as,
\begin{align}
\label{ent1}
S_2=\pi{r_2}^2\,.
\end{align}
We plot Eq.~(\ref{ent1}) in FIG.~\ref{Fig:2}~\subref{fig:2b}.
As this figure show, we have always positive entropy.
The Hawking temperatures associated with the BH solution (\ref{W6}) is plotted in FIG.~\ref{Fig:2}~\subref{fig:2c}.
From this figure, one can show that we have always a positive temperature for $r_2$.
%\begin{figure}
%\centering
%\subfigure[~The Hawking temperature of the black hole solution (\ref{sol}) ]{\label{fig:3a}\includegraphics[scale=0.4]{JMThr4}}\hspace{0.2cm}
%\subfigure[~The Hawking temperature of the black hole solution (\ref{sol1})]{\label{fig:3b}\includegraphics[scale=0.4]{JMTh33}}
%\caption{{\bf Schematic plot of the Hawking temperature of the two black holes (\ref{sol}) and (\ref{sol1}) versus the dimensional parameter $\alpha$ and $r_+$ , respectively.}}
%\label{Fig:3}
%\end{figure}
%From Eq.~(\ref{en}), the quasi-local energy of the black hole (\ref{sol1}) is plotted in FIG.~\ref{Fig:2}~\subref{fig:2e} and \ref{Fig:2}~\subref{fig:2f}. These figures show that we have always positive quasi-local energy for the black hole (\ref{sol1}).
%\begin{figure}
%\centering
%\subfigure[~The quasilocal energy of the black hole solution (\ref{sol}) ]{\label{fig:4a}\includegraphics[scale=0.4]{JMThr3E}}\hspace{0.2cm}
%\subfigure[~The quasilocal energy of the black hole solution (\ref{sol1})]{\label{fig:4b}\includegraphics[scale=0.4]{JMTh3E1}}
%\caption{\bf {Schematic plot of the quasilocal energy of the black holes (\ref{sol}) and (\ref{sol1}) versus the dimensional parameter $\alpha$ and $r_+$, respectively.}}
%\label{Fig:4}
%\end{figure}
To investigate the thermodynamical stability of BHs, the formula of the heat capacity $H\left( r_2 \right)$
at the event horizon should be derived.
The heat capacity is defined as follows \cite{Nouicer:2007pu,DK11,Chamblin:1999tk},
\begin{align}
\label{heat-capacity11}
H_c\equiv H \left( r_2 \right)=\frac{\partial M_2}{\partial T_2}=\frac{\partial M_2}{\partial r_2} \left(\frac{\partial T_2}{\partial r_2}\right)^{-1}\, .
\end{align}
The BH will be thermodynamically stable if its heat capacity $H_c$ is positive.
On the other hand, it will be unstable if $H_c$ is negative.
{ As well-known, the heat capacity of the Schwarzschild black hole in GR is negative and therefore the solution is unstabel,
which corresponds to the Hawking evaporation.}
Substituting (\ref{m2}) and (\ref{temp11}) into (\ref{heat-capacity11}), we obtain the heat capacity as follows,
% The heat capacity of the BH solution (\ref{W6}) takes the form:
\begin{align}
\label{heat-capacity111_0}
H_c=\frac{\pi {r_2}^2 \left( c_1{r_2}^4-{r_2}^6+5c_2 \right)}{2M{r_2}^5-3c_1{r_2}^4-21c_2}\, .
\end{align}
The free energy in the grand canonical ensemble, which is calledthe  Gibbs free energy, can be defined as \cite{Zheng:2018fyn,Kim:2012cma},
\begin{align}
\label{enr1}
G\left( r_2 \right)=E\left( r_2 \right)-T\left( r_2 \right)S\left( r_2 \right)%+P(r_+)V(r_+),
\end{align}
%$V$ is the geometric volume of the black hole, $P$ is the pressure which is represented by the radial equation of (\ref{f1}), i.e. $I_r{}^r$,
where $E\left( r_2 \right)$, $T\left( r_2 \right)$, and $S\left( r_2 \right)$ are the quasi-local energy, the temperature and entropy at the event horizons, respectively.
Substituting Eqs.~(\ref{m2}), (\ref{temp11}), and (\ref{ent}) into (\ref{enr1}), we obtain the Gibbs free energy of the BH (\ref{W6}) in the following form,
\begin{align}
\label{enr}
G\left( r_2 \right)=\frac{{r_2}^6+3c_1{r_2}^4+7c_2}{4{r_2}^5}\,.
\end{align}
We plot the behavior of the Gibbs free energy in FIG.~\ref{Fig:2}~\subref{fig:2e}, which shows that the BH solution (\ref{W6}) with $r_2$ is unstable.
%\begin{figure}
%\centering
%\subfigure[~The free energy of the black hole solution (\ref{sol1}) vs. the mass $m$ using the horizon $r_+$ %]{\label{fig:3a}\includegraphics[scale=0.25]{JRKIGibm}}\hspace{0.2cm}
%\subfigure[~The free energy of the black hole solution (\ref{sol1})vs. the mass $m$ using the horizon %$r_-$]{\label{fig:3b}\includegraphics[scale=0.25]{JRKIGibm1}}\hspace{0.2cm}
%\subfigure[~The free energy of the black hole solution (\ref{sol}) vs. the parameter $c_0$ using the horizon $r_+$ ]{\label{fig:3c}\includegraphics[scale=0.25]{JRKIGibc}}\hspace{0.2cm}
%\subfigure[~The free energy of the black hole solution (\ref{sol1}) vs. the parameter $c_0$ using the horizon $r_-$ ]{\label{fig:3d}\includegraphics[scale=0.25]{JRKIGibc1}}
%\caption{{Schematic plot of the free energy of the black holes (\ref{sol1}) vs. the mass $m$ and the parameter $c_0$ respectively.}}
%\label{Fig:3}
%\end{figure}
%The behaviors of the Gibbs energy of our black hole that is presented in figure \ref{Fig:2}\subref{fig:2e}, for particular values of the model parameters show the Gibbs energy is negative which means that we have a stable black hole.

\subsection{Thermodynamics of the BH (\ref{W66}, \ref{W77})}

\begin{figure}
\centering
\subfigure[~The plot of the function $f(r)$ vs. the radial coordinate $r$ for the BH (\ref{W66})]{\label{fig:3a}\includegraphics[scale=0.25]{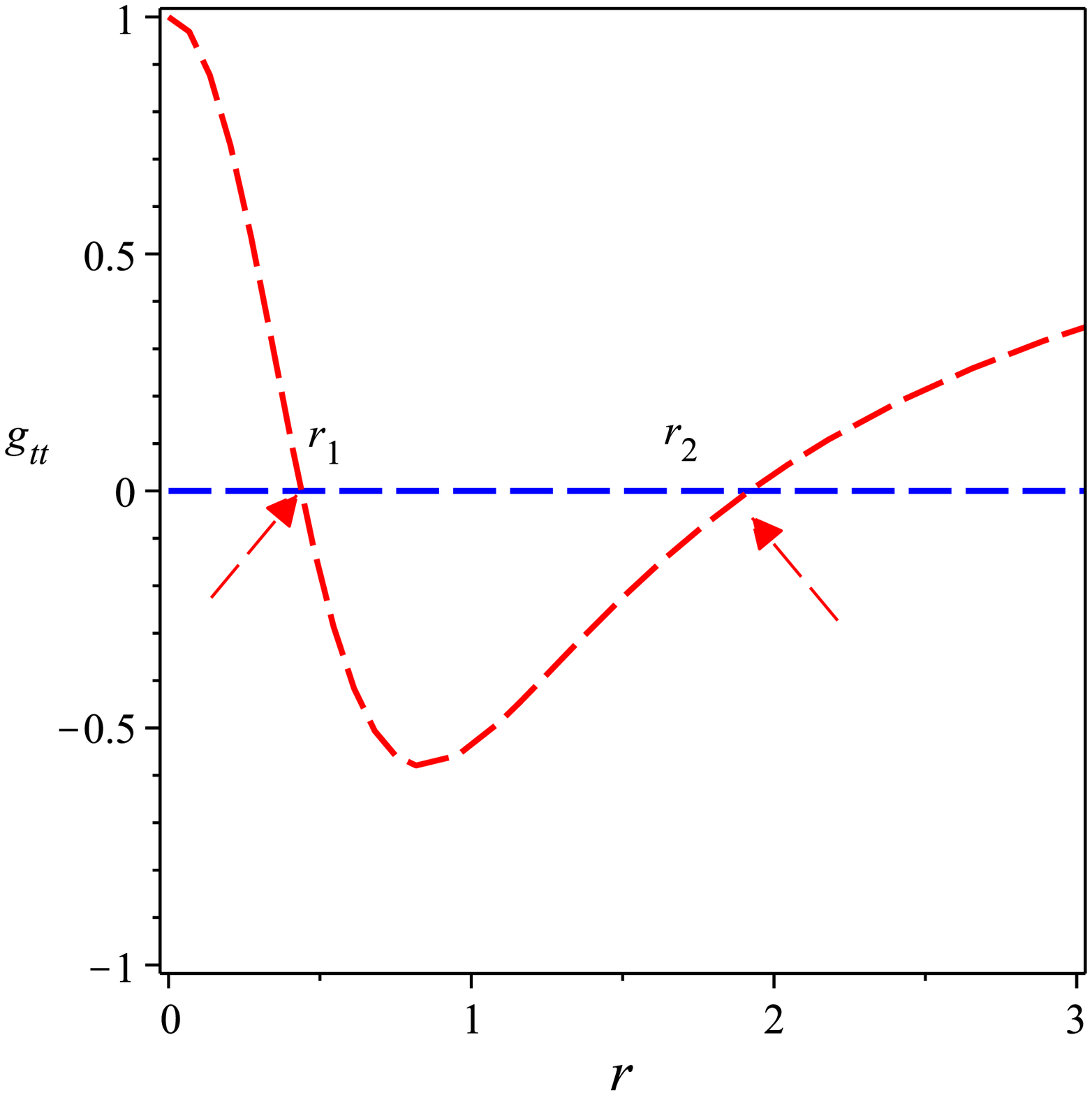}}\hspace{0.2cm}
\subfigure[~The plot of the function $f(r)$ and $f_1(r)$ vs. the radial coordinate $r$ for the BH (\ref{W66}) ]{\label{fig:3b}\includegraphics[scale=0.25]{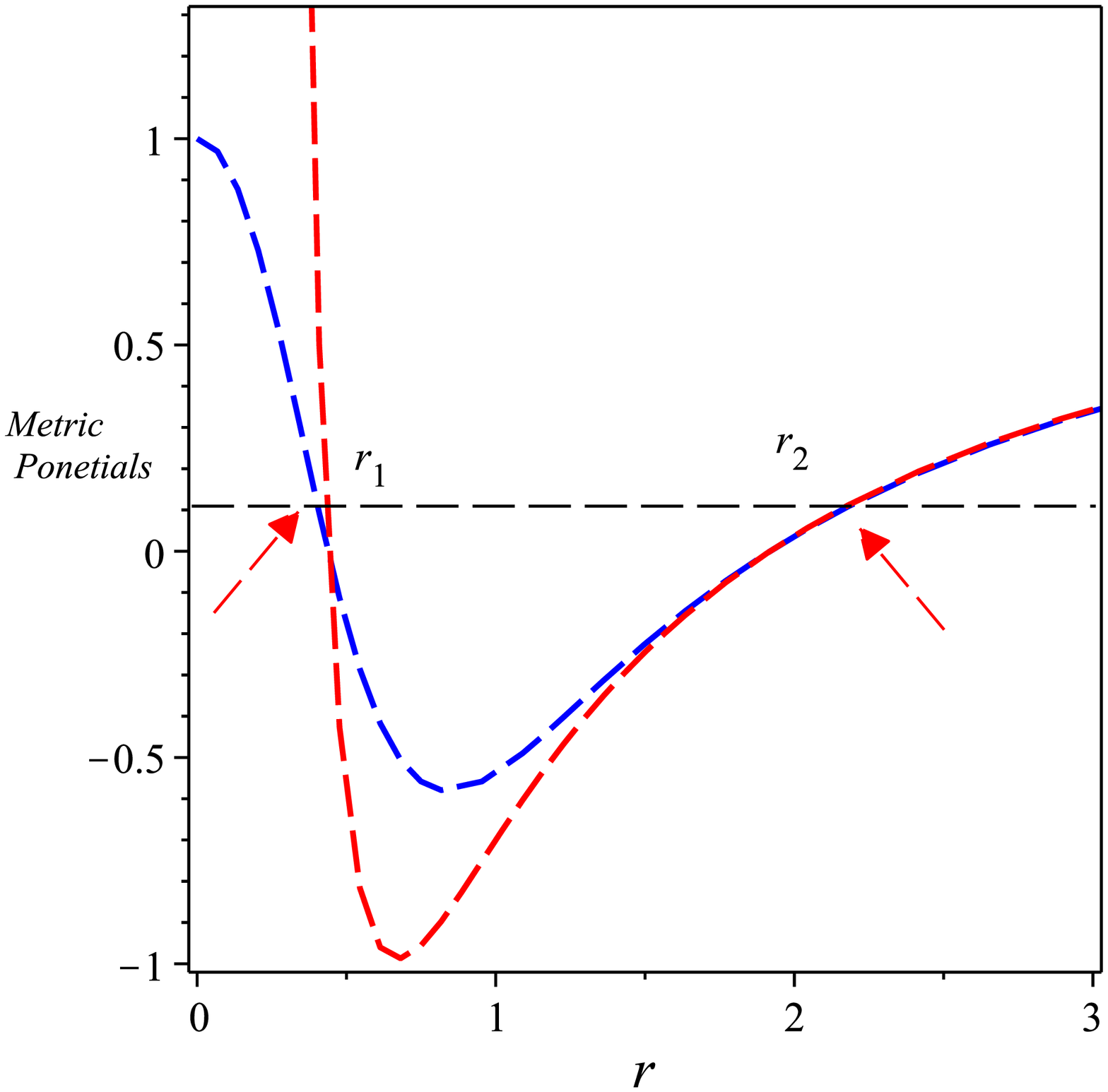}}\hspace{0.2cm}
\subfigure[~The plot of GB invariant given by Eq.~(\ref{G33}) vs. the radial coordinate $r$ for the BH (\ref{W66})]{\label{fig:3c}\includegraphics[scale=0.25]{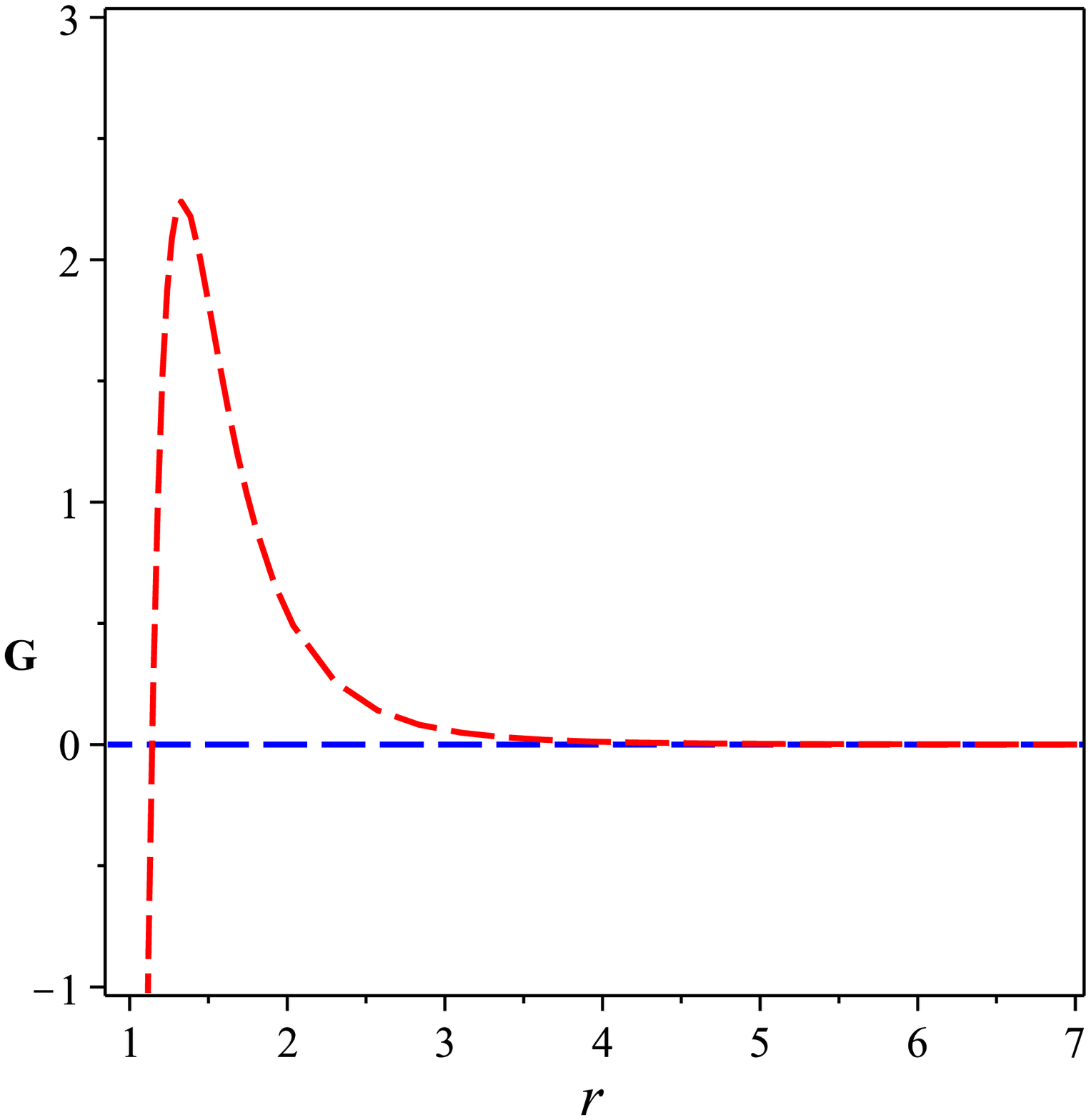}}\hspace{0.2cm}
\subfigure[~The function $h(\chi)$ vs. $\chi$ for the BH (\ref{W77})]{\label{fig:3d}\includegraphics[scale=0.25]{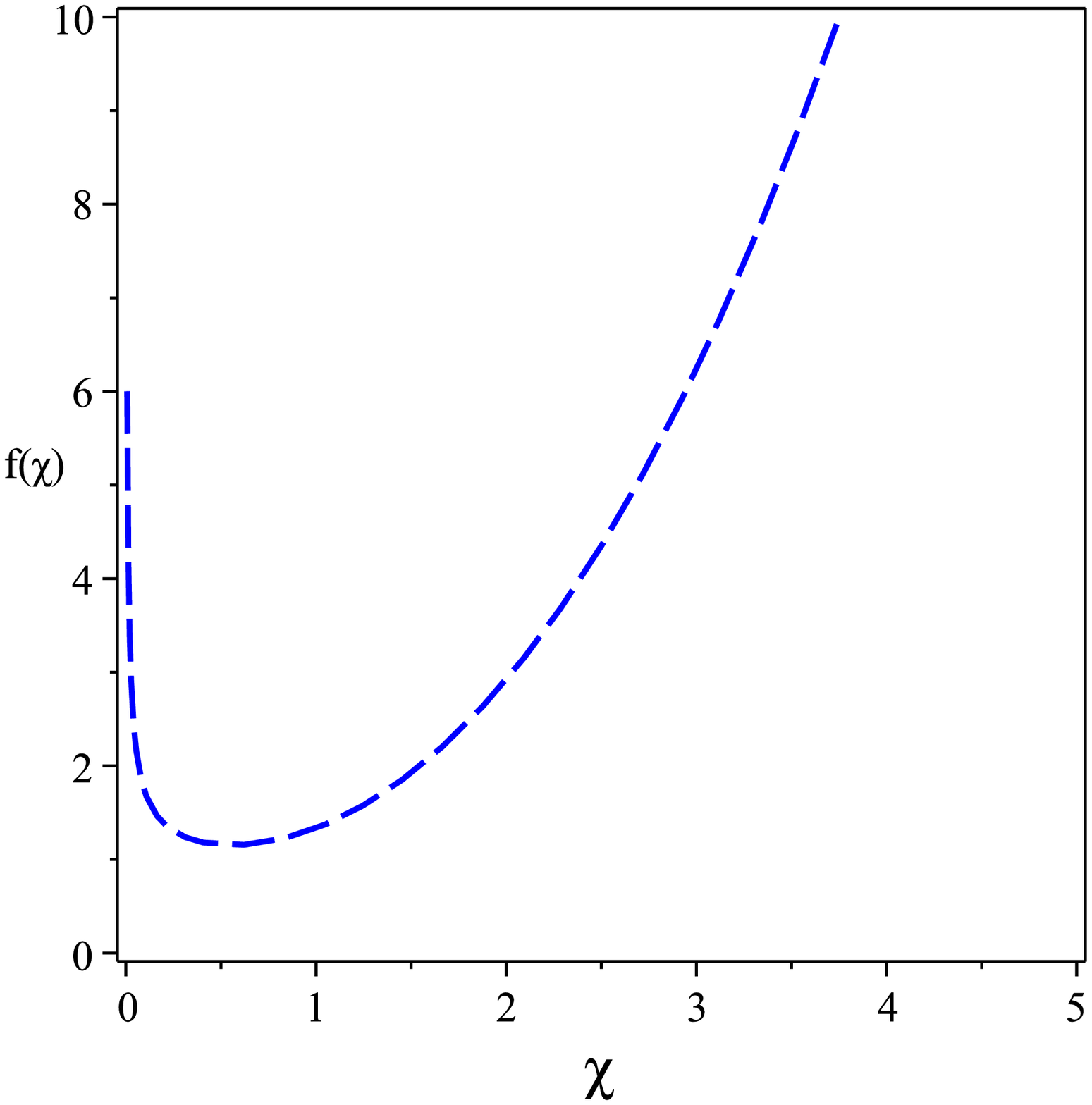}}\hspace{0.2cm}
\subfigure[~The potential $V(\chi)$ vs. $\chi$ for the BH (\ref{W77}) ]{\label{fig:3e}\includegraphics[scale=0.25]{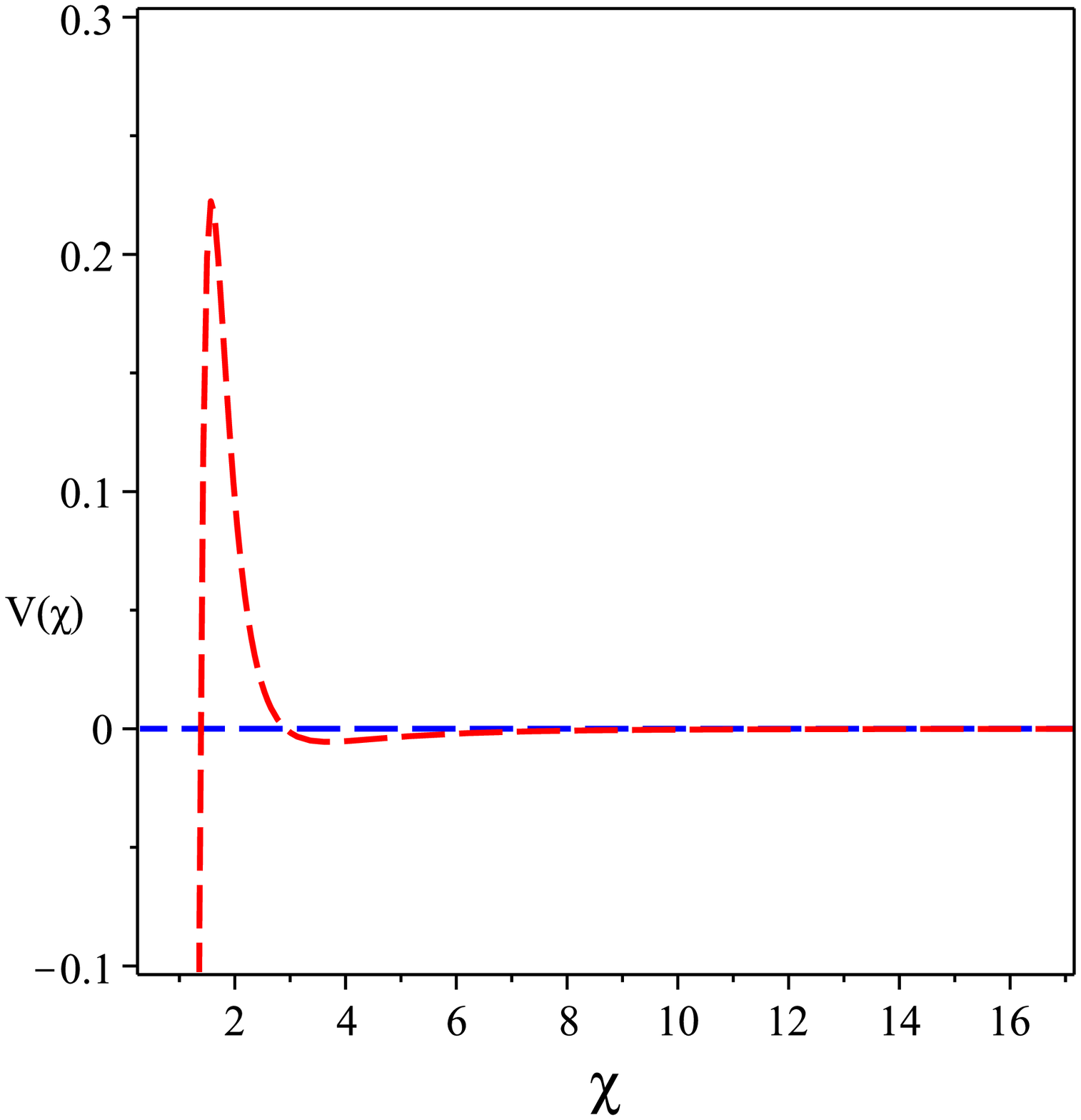}}
\subfigure[~The Lagrangian potential $\lambda$ vs. $r$ for the BH (\ref{W77})]{\label{fig:3f}\includegraphics[scale=0.25]{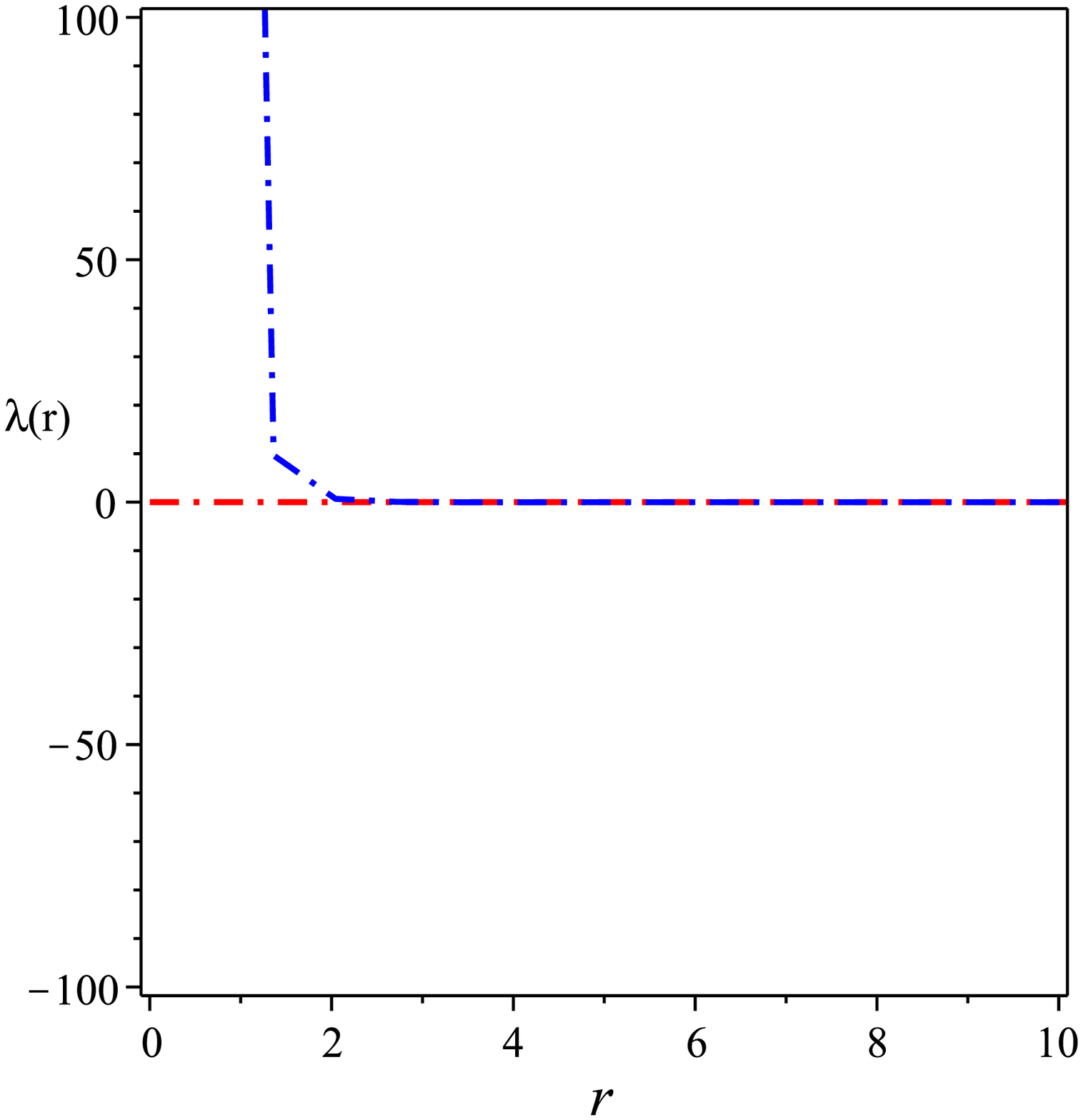}}
\caption{Schematic plots of the radial coordinate $r$   \subref{fig:3a} vs. the function $f$;   \subref{fig:3b} vs. the metric potentials,   \subref{fig:3c}
vs. the GB invariant given by Eq.~(\ref{W88}); \subref{fig:3d} the function $h$   vs. $\chi$; \subref{fig:3e} the potential $V$  vs.  $\chi$, and \subref{fig:3f} the Lagrange multiplier $\lambda(r)$ vs. $r$.}
\label{Fig:3}
\end{figure}

\begin{figure}
\centering
\subfigure[~{ The horizons, $r_1$ and $r_2$, of the BH solution (\ref{W66})}]{\label{fig:4a}\includegraphics[scale=0.25]{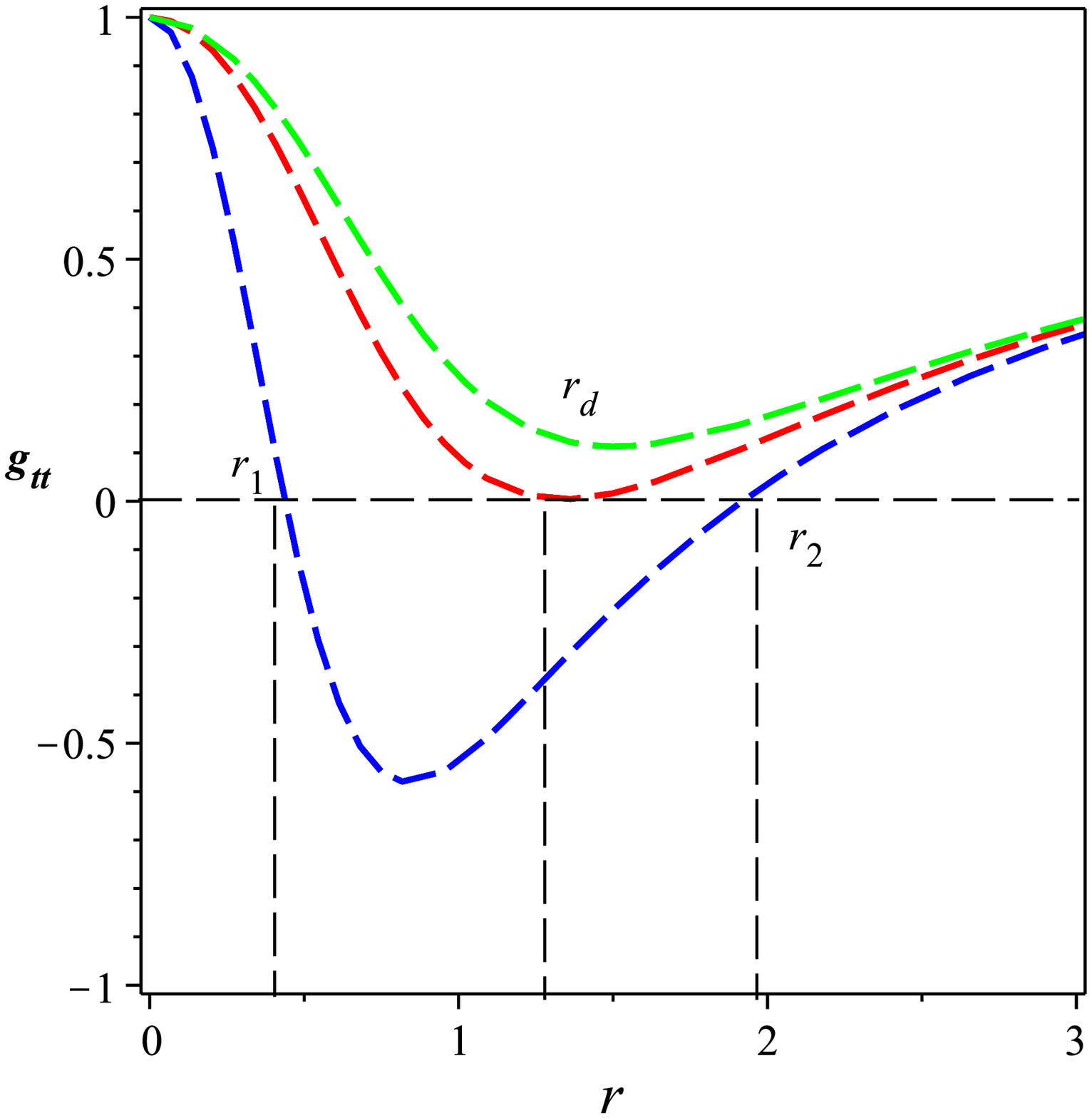}}\hspace{0.2cm}
%\subfigure[~The entropy of the black hole solution (\ref{W6}) vs. $r$ ]{\label{fig:2b}\includegraphics[scale=0.25]{JFRMMM_PKY_ent}}\hspace{0.2cm}
\subfigure[~The temperature of the BH solution (\ref{W66}) vs. $r$ ]{\label{fig:4b}\includegraphics[scale=0.25]{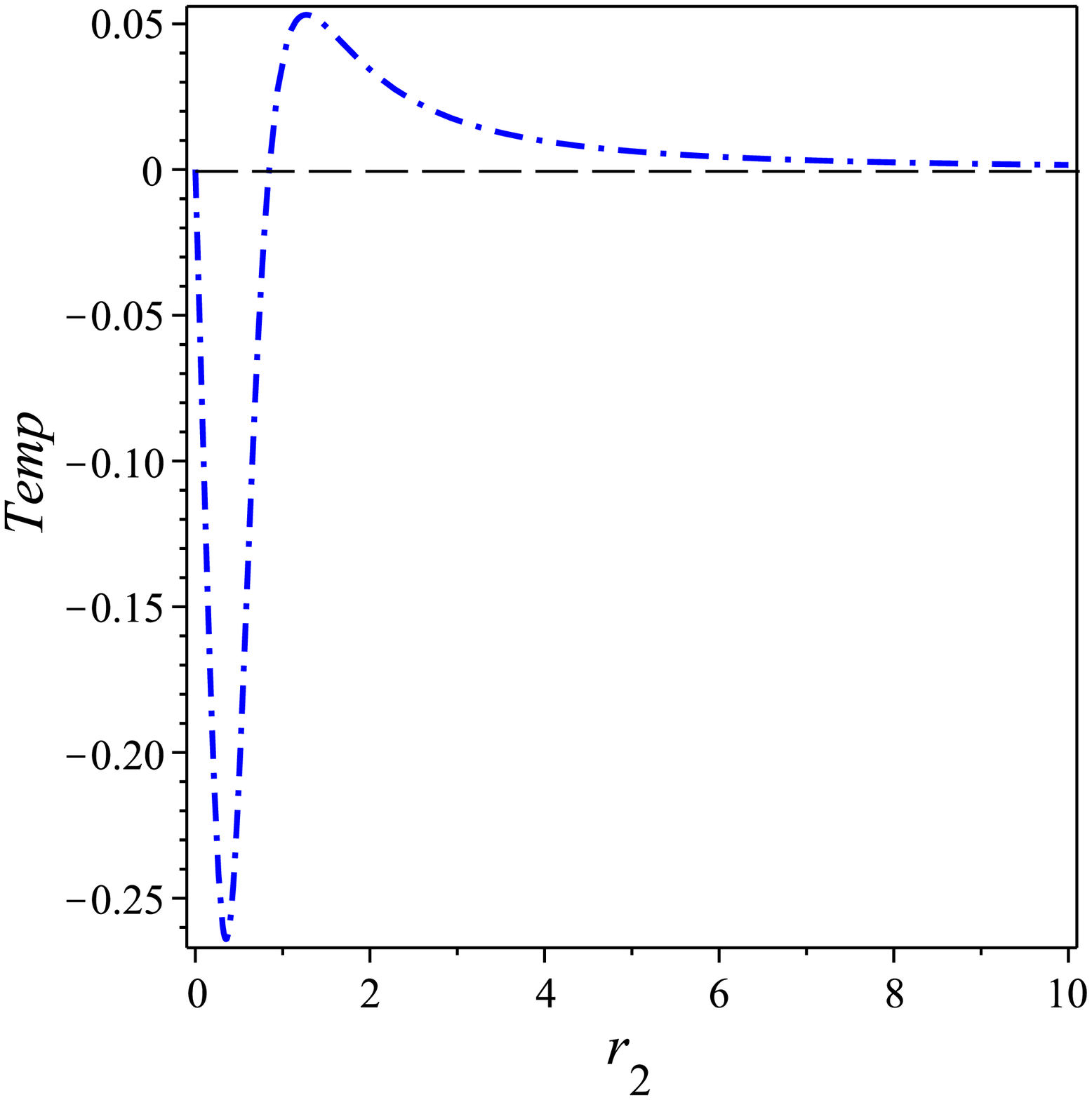}}\hspace{0.2cm}
\subfigure[~The heat capacity of the BH solution (\ref{W66}) vs. $r$]{\label{fig:4c}\includegraphics[scale=0.25]{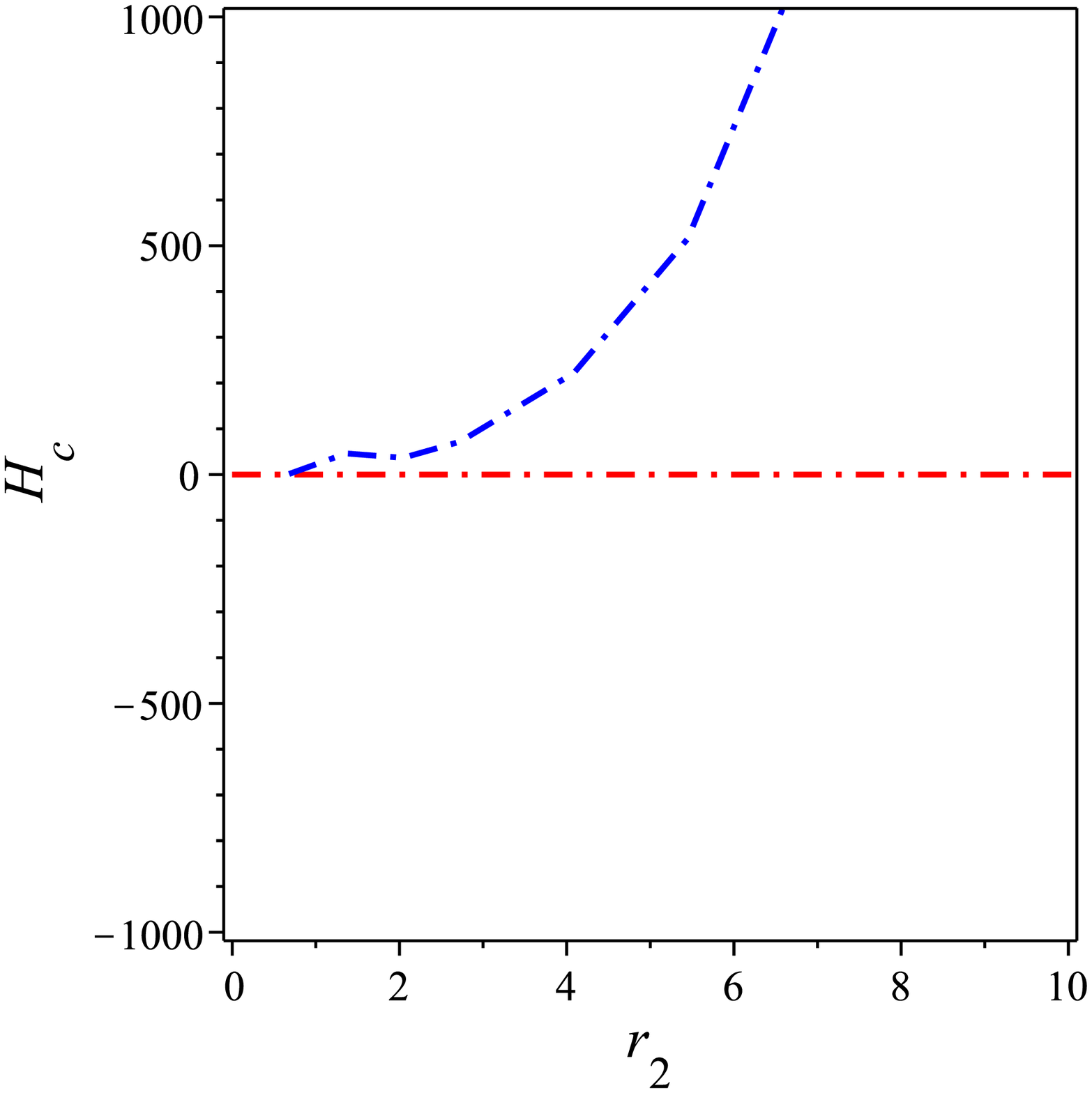}}
\subfigure[~{ The Gibbs free energy of the BH solution (\ref{W66}) vs. $r$}]{\label{fig:4d}\includegraphics[scale=0.25]{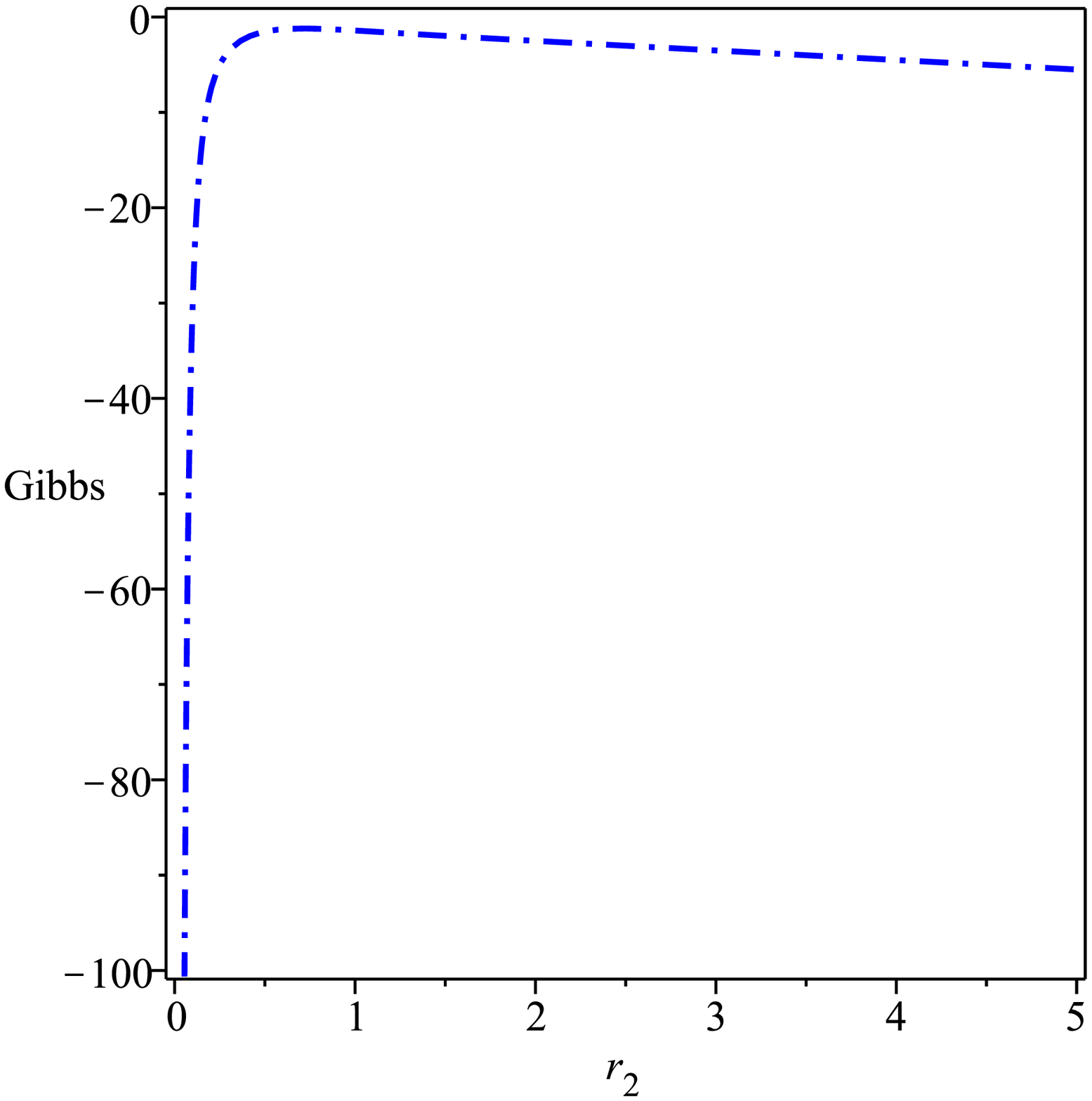}}
\caption{{ {Schematic plot of \subref{fig:4a} the horizons,$r_1$ and $r_2$, of the BH solution (\ref{W66}); \subref{fig:2b}
the Hawking temperature of the BH solution (\ref{W66}); \subref{fig:4c} the heat capacity of the BH solution (\ref{W6}) finally, \subref{fig:4d} the Gibbs free energy of the BH solution (\ref{W66})}.}}
\label{Fig:4}
\end{figure}

In this section, we will study the thermodynamics of the BH solution in (\ref{W66}, \ref{W77}).
For this aim, by assuming $r$ is large, we rewrite the metric as follows,
\begin{align}
\label{mpab111}
f(r)=1+\frac{\alpha r^2}{r^3+\beta}\approx 1-\frac{2M}{r}+\frac{2M\beta}{r^4}-\frac{2M\beta^2}{r^7}\,, \quad \mbox{where} \quad \alpha=-2M\,.
\end{align}
By using Eqs.~(\ref{mpab111}) and (\ref{W66}) we obtain
\begin{align}
\label{metaf11}
ds^2=-\left[1-\frac{2M}{r}+\frac{2M\beta}{r^4}-\frac{2M\beta^2}{r^7}\right]dt^2 +\frac{dr^2}{1-\frac{2M}{r}+\frac{\beta}{r^3}}
+r^2 \left( d\theta^2+\sin^2\theta d\phi^2 \right)\,,
\end{align}
which is asymptotically approaches flat space-time but is not equal to the Schwarzschild space-time due to the contribution of the extra term including $\beta$.
It is easy to check that when the term $\beta$ vanishes, the geometry reduces to the Schwarzschild space-time.
 From Eq.~(\ref{metaf11}), we obtain an expression of $M$ in terms of the radial coordinate $r$ as follows,
\begin{align}
\label{m222}
M=\frac{r}{2}\left(1+\frac{2M\beta}{r^4}-\frac{2M\beta^2}{r^7}\right)\, .
\end{align}
The metric of the line-element given by (\ref{metaf11}) has two horizons as shown in FIG.~\ref{Fig:3}~\subref{fig:3a} shows.
These two horizons are created by the constants $M$ and $\beta$.
When $\beta$ vanishes, the geometry of the Schwarzschild geometry is reproduced.
The behavior of the metric is drawn in FIG.~\ref{Fig:3}~\subref{fig:3b}, which shows clearly that there are two horizons related to the BH solution (\ref{W66}).
As Eq.~(\ref{G33}) shows, the BH solution (\ref{metaf11}) has a non-trivial expression of the GB invriant, whose behavior is shown in FIG.~\ref{Fig:3}~\subref{fig:3c}.
The behavior of the physical quantities related to the BH solution (\ref{W77}) like $h(\chi)$, $V(\chi)$, and the Lagrange multiplier field $\lambda$
are shown in FIG.~\ref{Fig:3}~\subref{fig:3d}, \ref{Fig:3}~\subref{fig:3e}, and \ref{Fig:3}~\subref{fig:3f}.
To show how many horizons appear in the BH solution of Eq.~(\ref{W66}), we plot the metric $g_{00}$ in FIG.~\ref{Fig:4}~\subref{fig:4a}.
As FIG.~\ref{Fig:4}~\subref{fig:4b} shows that in the case $\beta=0.3$ \footnote{{In this study we use Eq. \eqref{mpab111} and put $\alpha=-2$ which yields $M=1$.}}, we have two horizons and when $\beta=1.3$, we have no horizon.
Also in FIG.~\ref{Fig:4}~\subref{fig:4b}, we show the region where the black hole has naked singularity, i.e., when $\beta>1.3$.

By using Eq.~(\ref{temp}), we obtain the Hawking temperature of the BH solution (\ref{W66}) in the following form,
\begin{align}
\label{temp2}
T_2 =\frac{\alpha r_2 \left(2\beta-{r_2}^3 \right)}{4\pi \left( {r_2}^4+\beta \right)^2}\,,
\end{align}
We show the behavior of Eq.~(\ref{temp2}) in FIG.~\ref{Fig:4}~\subref{fig:4b}.
{ From this figure one can show that we have a positive temperature for $r_2>r_d$ and negative temperature for $r_2<r_d$ where $r_d$ is the degenerate horizon as shown in FIG.~\ref{Fig:4}~\subref{fig:4a}.}
%\begin{figure}
%\centering
%\subfigure[~The Hawking temperature of the black hole solution (\ref{sol}) ]{\label{fig:3a}\includegraphics[scale=0.4]{JMThr4}}\hspace{0.2cm}
%\subfigure[~The Hawking temperature of the black hole solution (\ref{sol1})]{\label{fig:3b}\includegraphics[scale=0.4]{JMTh33}}
%\caption{{\bf Schematic plot of the Hawking temperature of the two black holes (\ref{sol}) and (\ref{sol1}) versus the dimensional parameter $\alpha$ and $r_+$ , respectively.}}
%\label{Fig:3}
%\end{figure}
%From Eq.~(\ref{en}), the quasi-local energy of the black hole (\ref{sol1}) is plotted in FIG.~\ref{Fig:2}~\subref{fig:2e} and \ref{Fig:2}~\subref{fig:2f}. These figures show that we have always positive quasi-local energy for the black hole (\ref{sol1}).
%\begin{figure}
%\centering
%\subfigure[~The quasilocal energy of the black hole solution (\ref{sol}) ]{\label{fig:4a}\includegraphics[scale=0.4]{JMThr3E}}\hspace{0.2cm}
%\subfigure[~The quasilocal energy of the black hole solution (\ref{sol1})]{\label{fig:4b}\includegraphics[scale=0.4]{JMTh3E1}}
%\caption{\bf {Schematic plot of the quasilocal energy of the black holes (\ref{sol}) and (\ref{sol1}) versus the dimensional parameter $\alpha$ and $r_+$, respectively.}}
%\label{Fig:4}
%\end{figure}
Substituting (\ref{m222}) and (\ref{temp2}) into (\ref{heat-capacity11}), we obtain the heat capacity as follows,
\begin{align}
\label{heat-capacity111}
H_c=\frac{2\pi \left( \beta- {r_2}^3 \right) \left({r_2}^3+\beta \right)}{{r_2}^3\alpha \left({r_2}^6-7\beta {r_2}^3+\beta^2 \right)}\, .
\end{align}
We show the behavior of Eq.~(\ref{heat-capacity111}) in FIG.~\ref{Fig:4}~\subref{fig:4c}.
By substituting Eqs.~(\ref{ent}), (\ref{m222}), and (\ref{temp2}) into (\ref{enr}), we obtain the Gibbs free energy of the BH (\ref{W66}) in the following form,
\begin{align}
\label{enrB}
G \left( r_+ \right)=-\frac{4{r_2}^9+12\beta {r_2}^6+12\beta^2 {r_2}^3+4\beta^3-\alpha {r_2}^8+2\alpha \beta {r_2}^5}{4{r_2}^2({r_2}^3+\beta)^2}\,.
\end{align}
We plot the behavior of the Gibbs free energy in FIG.~\ref{Fig:4}~\subref{fig:4d}, which shows that the BH solution (\ref{W66}) with $r_2$ is unstable.
%\begin{figure}
%\centering
%\subfigure[~The free energy of the black hole solution (\ref{sol1}) vs. the mass $m$ using the horizon $r_+$ %]{\label{fig:3a}\includegraphics[scale=0.25]{JRKIGibm}}\hspace{0.2cm}
%\subfigure[~The free energy of the black hole solution (\ref{sol1})vs. the mass $m$ using the horizon %$r_-$]{\label{fig:3b}\includegraphics[scale=0.25]{JRKIGibm1}}\hspace{0.2cm}
%\subfigure[~The free energy of the black hole solution (\ref{sol}) vs. the parameter $c_0$ using the horizon $r_+$ ]{\label{fig:3c}\includegraphics[scale=0.25]{JRKIGibc}}\hspace{0.2cm}
%\subfigure[~The free energy of the black hole solution (\ref{sol1}) vs. the parameter $c_0$ using the horizon $r_-$ ]{\label{fig:3d}\includegraphics[scale=0.25]{JRKIGibc1}}
%\caption{{Schematic plot of the free energy of the black holes (\ref{sol1}) vs. the mass $m$ and the parameter $c_0$ respectively.}}
%\label{Fig:3}
%\end{figure}
%The behaviors of the Gibbs energy of the BH that is presented in figure \ref{Fig:3}\subref{fig:3e}, for particular values of the model parameters show the Gibbs energy is negative which means that we have a stable black hole.
%%%%%%%%%%%%%%%%%%%%%%%%%%%%%%%%%%%%%%%%%%%%%%%%%%%%%%%%%%%%%%%%%%%%%%%%%%%%%%%%%%%%%%%%%%%%%%%%%%%%%

\section{Motion of Particle}\label{sec:pheno}

To show the effect of modified GB theory on observables, we study the motion of a test particle in the background solution given by the metric (\ref{W6}) and (\ref{W66}).
We consider the photon sphere around the BH and the perihelion shift of circular orbits.
For the time being, the photon sphere becomes of particular interest because it explains the edge of the shadow of a BH while the perihelion shift
was already derived in \cite{Bahamonde:2019zea,DeBenedictis:2016aze}.

\subsection{Geodesic equation and effective potential}

In this subsection, we study the geodesic equation in the space-time given by Eq.~(\ref{W6}).
For this aim, we define the worldline $q(\tau)$ of a test particle in a curved space-time by the Euler-Lagrange equations which is defined by,
\begin{align}
\label{4}
\frac{d}{d\tau}\left(\frac{\partial \mathcal{L}}{\partial \dot q^\mu}\right)-\frac{\partial \mathcal{L}}{\partial q^\mu}=0\,,
\end{align}
for the Lagrangian
\begin{align}
\label{7}
2\mathcal{L}=g_{\mu\nu}\dot q^{\mu}\dot q^\nu=f(r)\dot{t}^2-\frac{\dot r^2}{f(r)}-r^2\dot{\theta}^2-r^2\sin^2{\theta}\dot{\phi}^2\,,
\end{align}
with $q^{\mu}(\tau)= \left(t \left(\tau \right), r \left( \tau \right), \theta \left(\tau \right), \phi \left(\tau \right)\right)$
and $\dot q^\mu$ refers to the derivative of $q^\mu$ w.r.t. the affine parameter $\tau$.

%%%%%%
%%%%%%
%%%%%%%

We solve the Euler-Lagrange equations (\ref{4}) in the spherically symmetric space-times and we focus on the motion of the equatorial plane with $\theta=\pi/2$.
under the assumption, we obtain the conserved quantities, i.e., the energy $E$ and angular momentum $L$ as follows,
\begin{align}
\label{E}
E =&\, \frac{\partial \mathcal{L}}{\partial \dot t} = f(r)\dot{t}=\left(1-\frac{2M}{r}+\frac{c_1}{r^2}+\frac{c_2}{r^6}\right)\dot{t}\,,\\
\label{L}
L =&\, \frac{\partial \mathcal{L}}{\partial \dot \phi} = r^2 \dot \phi\,.
\end{align}
Using the above conserved quantities (\ref{E}) and (\ref{L}), we obtain the effective potential in classical mechanics.
Because $2\mathcal{L}=0$ for the massless particle and $2\mathcal{L}=1$ for the massive particle, by deleting
$\dot t$ and $\dot \phi$ by using Eqs.~ (\ref{E}) and (\ref{L}), and by putting $\theta=\pi/2$ (constant), we obtain
\begin{align}
\label{177}
\frac{E^2}{1-\frac{2M}{r}+\frac{c_1}{r^2}+\frac{c_2}{r^6}}-\frac{\dot r^2}{1-\frac{2M}{r}+\frac{c_1}{r^2}+\frac{c_2}{r^6}} - \frac{L^2}{r^2} = \sigma\,,
\end{align}
where $\sigma = 0$ for massless particles and $\sigma = 1$.
We rewrite Eq.~(\ref{177}) as follows,
\begin{align}
\label{cons1}
0 = \frac{1}{2}\dot r^2 - \frac{1}{2} E^2 + \frac{1}{2} \frac{L^2}{r^2}\left(1-\frac{2M}{r}+\frac{c_1}{r^2}+\frac{c_2}{r^6}\right) + \frac{1}{2} \sigma \left(1-\frac{2M}{r}+\frac{c_1}{r^2}+\frac{c_2}{r^6}\right) \,,
\end{align}
from which we can read off the effective potential $\mathcal{V}(r)$,
\begin{align}
\label{eq:pot11}
\mathcal{V}(r) =&\,  \frac{1 }{2}\left(1-\frac{2M}{r}+\frac{c_1}{r^2}+\frac{c_2}{r^6}\right) \left(\frac{L^2}{r^2}+ \sigma\right) - \frac{1}{2} E^2 \, ,
\end{align}
and we rewrite (\ref{cons1}) as follows,
\begin{align}
\label{EffE}
\frac{1}{2}\dot r^2 + \mathcal{V}(r)=0\,.
\end{align}
For the study of the perihelion shift, we reparametrize $r(\tau)$ as $r(\phi)$, which yields,
\begin{align}
\label{eq:rphi1}
\frac{1}{2}\frac{\dot r^2}{\dot \phi^2} + \frac{1}{\dot \phi^2}\mathcal{V}(r) = \frac{1}{2}\left(\frac{dr}{d\phi}\right)^2 + \frac{r^4}{L^2}\mathcal{V}(r)=0\,.
\end{align}

\subsection{Photon sphere and perihelion shift of space-time \eqref{W6}}\label{ssec:phsph}

For a circular orbit where $r=\mbox{const.}$, $\dot r=0$, the effective potential and its derivative have to vanish i.e., 
we have to solve both equations $\mathcal{V} = 0$ and $\mathcal{V}' = 0$.  

When $c_2=\sigma=0$, the effective potential $\mathcal{V}(r)$ reduces to, 
\begin{align}
\label{eq:pot11B}
\mathcal{V}(r) = \frac{1 }{2}\left(1-\frac{2M}{r}+\frac{c_1}{r^2} \right) \frac{L^2}{r^2} - \frac{1}{2} E^2 \, .
\end{align}
When $r$ is large, $\mathcal{V}(r)$ is monotonically decreasing function of $r$.
On the other hand, when $r$ is small, $\mathcal{V}(r)$ behaves as $\mathcal{V}(r) \sim \frac{c_1 L^2}{2 r^4}$ and
therefore if $c_1>0$,  $\mathcal{V}(r)$ goes to positive infinity and  if $c_1<0$,  $\mathcal{V}(r)$ goes to negative infinity.

For circular photon orbits, by solving the equations $\mathcal{V}(r) = \mathcal{V}'(r) =0$ for the potential $\mathcal{V}(r)$ 
in Eq.~(\ref{eq:pot11B}),  we find, 
\begin{align}
\label{p133}
r=&\, \frac{3}{2}M\pm\frac{1}{2}\sqrt{9M^2-8c_1}	\,, \nonumber\\
L_{\pm}=&\, \pm \frac{(3M+\sqrt{9 M^2 - 8 c_1})^2 E}{2 \sqrt{6M^2+2M\sqrt{9 M^2-8 c_1}-4c_1}}\,,
%&L_{\pm}= \pm 1 / \sqrt{2 c1-2 M^2}E \sqrt{-8 c1^2 - 9 M^3 \left(\sqrt{9 M^2-8
%c1}+3 M\right)+4 c1 M \left(2 \sqrt{9 M^2 - 8 c1}+9 M\right)}\,.
\end{align}
where the value of $r$ given in the first equation of Eq.~(\ref{p133}) is used in the second equation of (\ref{p133}). 
Eq.~(\ref{p133}) gives the value of the Schwarzschild when $c_1=0$, i.e., $r=3M$ and $L_{\pm}=3\sqrt{3}M\,E$.

The expression of $r$ in (\ref{p133}) tells that when $c_1<0$, there is only one extremum $r= \frac{3}{2}M + \frac{1}{2}\sqrt{9M^2-8c_1}$.
The behavior if the potential tells that the extremum is a maximum and therefore the orbit $r$ is unstable.
On the other hand, when $\frac{9 M^2}{8}> c_1>0$, there are two extrema $r= \frac{3}{2}M\pm\frac{1}{2}\sqrt{9M^2-8c_1}$.
The behavior of the potential tells that the larger extremum $r= \frac{3}{2}M + \frac{1}{2}\sqrt{9M^2-8c_1}$ is a local maximum and therefore the
orbit corresponding to the extremum is unstable but the smaller extremum $r= \frac{3}{2}M - \frac{1}{2}\sqrt{9M^2-8c_1}$ is a local minimum 
and therefore the orbit corresponding to the extremum is stable.

%Moreover, Eq.~(\ref{p133}) tells $c_1<\frac{9}{8}M^2$, that is, there is a lower limit for the mass $M$ in order that the circular orbit of the photon exists.

For circular timelike orbits $\sigma = 1$ for a massive particle, it is also possible to solve the equations $\mathcal{V} = 0$ and $\mathcal{V}' = 0$.
The obtained expressions are, however, not so insightful.
We consider a perturbation around a circular orbit $r= r_\mathrm{crc.}$ and by plugging in the ansatz $r(\phi)= r_\mathrm{crc.} + r_\phi(\phi)$ for (\ref{eq:rphi1}), we obtain,
\begin{align}
\left(\frac{d r_\phi}{d\phi}\right)^2 = - 2 \frac{(r_\mathrm{crc.} + r_\phi)^4}{h^2} \mathcal{V}\left( r_\mathrm{crc.} + r_\phi \right)\,.
\end{align}
Assuming that the ratio $r_\phi/r_c$ is small, the right-hand side can be expanded into powers of this parameter to second order
\begin{align}
\left(\frac{d r_\phi}{d\phi}\right)^2
 = - \frac{r_\mathrm{crc.}^4}{h^2} \mathcal{V}'' \left( r_\mathrm{crc.} \right){r_\phi}^2 + \mathcal{O}\left(\tfrac{{r_\phi}^3}{{r_\mathrm{crc.}}^3}\right) \,,
\end{align}
where we use the fact that $V \left(r_\mathrm{crc.} \right) = 0$ and $V' \left(r_\mathrm{crc.} \right)=0$ for circular orbits, as discussed above.
The above equation, which represents a simple harmonic oscillation, shows that the solution of $r_\phi$ oscillates with a wave number
$K = \sqrt{\frac{r_c^4}{h^2} \mathcal{V}'' \left(r_\mathrm{crc.} \right)}$ and thus the perihelion shift is given as,
\begin{align}
\Delta \phi =2\pi\left(\frac{1}{K}-1\right) =2\pi \left(\frac{h}{{r_\mathrm{crc.}}^2\sqrt{\mathcal{V}''\left( r_\mathrm{crc.} \right)}}-1\right)\,.
\end{align}
Now, we derive the explicit form of the perihelion shift for massive objects where the potential $\mathcal{V}$ with $\sigma=1$.
%, see \eqref{eq:pot11}, its first derivative $\mathcal{V}'$ and its second derivative $\mathcal{V}''$.
We evaluate the equations $\mathcal{V} \left( r_\mathrm{crc.} \right) = 0$ and $\mathcal{V}' \left(r_\mathrm{crc.} \right) = 0$ with $L= L_0 + \epsilon\, L_1$ and $E = E_0 +\epsilon\, E_1$.
The zeroth order behaviors of these equations determine $L_0 \left(r_\mathrm{crc.} \right)$ and $E_0 \left(r_\mathrm{crc.} \right)$ as follows,
\begin{align}
L_{0\pm} =&\, \pm {\frac{\sqrt{M{r_\mathrm{crc.}}^5-c_1 r^4-3 c_2}r}{\sqrt{{r_\mathrm{crc.}}^6
 -3 M{r_\mathrm{crc.}}^5+2c_1{r_\mathrm{crc.}}^4+4c_2}}}\,, \nonumber \\
E_{0\pm}=&\, \pm {\frac{\sqrt{4 M^2{r_\mathrm{crc.}}^5-2c_1 {r_\mathrm{crc.}}^4M-4 {r_\mathrm{crc.}}^6M+{r_\mathrm{crc.}}^7
+c_2 r_\mathrm{crc.}+c_1 {r_\mathrm{crc.}}^5-2 c_2 M}}{\sqrt{r_\mathrm{crc.}-3\,M}{r_\mathrm{crc.}}^{3}}}\,.
\end{align}
Having obtained the constants of motion for the circular orbit, we derive the perihelion shift by plugging in the values into $\mathcal{V}''(r_\mathrm{crc.}, L_0, E_0)$
to obtain $\mathcal{V}'' \left(r_\mathrm{crc.} \right)$ alone.
Due to the different solutions for the constants of motion, there exist two options to derive the perihelion shift
\begin{align}
\Delta\phi \left( L_{0+} \right)\,, \quad \Delta\phi \left( L_{0-} \right)\,,
\end{align}
which are related to each other through
\begin{align}
 \Delta\phi \left( L_{0-} \right) = - 4\pi - \Delta\phi \left(L_{0+} \right)\,.
\end{align}
By expanding the perihelion shift into a power series in the variables $q = \frac M{r_\mathrm{crc.}}$, $q_1 = \frac{c_1}{{r_\mathrm{crc.}}^2}$, and $q_2= \frac{c_2}{{r_\mathrm{crc.}}^6}$,
we obtain
\begin{align}
\label{eqn:perihelionshift1}
\Delta\phi \left(L_{0-} \right) =&\,  12 \left({\frac{\pi \left( \sqrt{q-q_1-3 q_2}+ \sqrt{q+12\,q_2} \right) }{ \left( q+12 q_2 \right)^{5/ 2}}}
 - {\frac{\pi}{ \left( q+12 q_2 \right)^2}} \right) {q_1}^4 \nonumber \\
&\, + \left\{\frac{18\pi \left( \sqrt{q-q_1-3 q_2}+\sqrt{(q+12 q_2)} \right) \left( 4q_2-3 q \right) }{\left( q+12 q_2 \right)^{5/2}}
%\right. \nonumber\\
%&\, \left. 
 -{\frac{3\pi \left(72 q - 96 q_2\right) }{4 \left( q+12 q_2 \right)^2}}\right\} {q_1}^3 \nonumber \\
&\, - \pi\left( \frac{4}{q+12 q_2} +{\frac{4 \left( \sqrt{q-q_1-3 q_2}
+\sqrt{q+12 q_2} \right) }{ \left( q+12 q_2 \right)^{3/2}}}+{\frac{3}{4 \left( q+12 q_2 \right)^2}} \left(8\left[ 24 {q_2}^2-qq_2+6q^2 \right]
%  \right. \nonumber\\
%& \left. 
+3  \left( 4q_2-3q \right)^2 \right) \right. \nonumber \\ 
&\, \left. +{\frac{3\pi \left( \sqrt{q-q_1-3 q_2}+\sqrt{q+12  q_2} \right) \left( 192 {q_2}^2-8 qq_2+48  q^2+ \left( 12 q_2-9 q \right)^2 \right) }
{ 4\left( q+12 q_2 \right)^{5/2} }} \right) {q_1}^2\nonumber\\
& + \left({\frac{ \pi \left( 12 q_2-9 q \right) }{-q-12 q_2}}-{\frac{3\pi \left( \sqrt{q-q_1-3 q_2}+\sqrt{q+12 q_2} \right) \left( 4q_2-3q \right) }{ \left( q+12 q_2 \right)^{3/2}}}
-{ \frac{9\pi \left( 24 {q_2}^2-qq_2+6 q^2 \right) \left( 4q_2-3q \right) }{ 2\left( -q-12 q_2 \right)^{ 2}}} \right.\nonumber\\
& \left. +{\frac{3\pi \left( \sqrt{q-q_1-3 q_2}+\sqrt{q+12 q_2} \right) \left( 24 {q_2}^2-qq_2+6 q^2 \right) \left( 12 q_2-9 q \right) }{2 \left( q+12 q_2 \right)^{5/2}}} 
\right) q_1 \nonumber \\
&\, +2 {\frac{2\pi \left( \sqrt{q -q_1-3 q_2}+\sqrt{q+12 q_2} \right) }{\sqrt{q+12 q_2}}} %\nonumber\\
%&
+{\frac{3\pi \left( \sqrt{q-q_1-3 q_2}+\sqrt{q+12 q_2} \right) \left( 24 {q_2}^2-qq_2+6 q^2 \right)^2}{ 4\left( q+12 q_2 \right)^{5/2}}} \nonumber \\
&\, -{\frac{3\pi \left( 24 {q_2}^2-qq_2+6 q^2 \right)^2}{4 \left( -q-12 q_2 \right)^2}} %\nonumber\\
%& 
 -{\frac{\pi \left( \sqrt{q-q_1-3 q_2}+\sqrt{q+12 q_2} \right) \left( 24 {q_2}^2-qq_2+6 q^2 \right) }{ \left( q+12 q_2 \right)^{3/2}}} \nonumber \\
&\, -{\frac{\pi \left( 24 {q_2}^2-qq_2+6 q^2 \right) }{q+12\,q_2}}+  \mathcal{O}\left( \left( qq_1q_2 \right)^3 \right) \,.
\end{align}
Eq.~(\ref{eqn:perihelionshift1}) when $q_1=q_2=0$, i.e., $c_1=c_2=0$, yields
\begin{align}
\label{eqn:perihelionshift}
\Delta\phi \left( L_{0-} \right) = 6 \pi q + 27 \pi q^2 + \mathcal{O} \left( q^3 \right) \,,
\end{align}
which coincides with the perihelion of the Schwarzschild solution.

The qualitative behavior of the perihelion shift is always the same, only the numerical values differ.
Eq.~(\ref{eqn:perihelionshift1}) shows that $q>q_1+3 q_2$, and the higher $q_1$ and $q_2$,
the smaller the influence of the perturbation and corrections to the perihelion shift appear only in higher orders in $q$.

Now we repeat the above perihelion of the BH solution (\ref{W6}) to the BH (\ref{W66}).

\subsection{Photon sphere and perihelion shift in space-time \eqref{W66}}\label{ssec:phsph6}

For the Lagrangian
\begin{align}
\label{77}
2\mathcal{L}=g_{\mu\nu}\dot q^{\mu}\dot q^{\nu}=f(r)\dot{t}^2-\frac{\dot r^2}{f_1(r)}-r^2\dot{\theta}^2-r^2\sin^2{\theta}\dot{\phi}^2\,,
\end{align}
with $q^{\mu}(\tau)=\left(t \left(\tau \right), r \left(\tau \right), \theta \left(\tau\right), \phi \left(\tau \right) \right)$,
and $\dot q^\mu$ refers to the derivative of $q^\mu$ w.r.t. the affine parameter $\tau$.

To solve the Euler-Lagrange equations, we apply the same procedure used above for the BH (\ref{W6}).
For the BH solution (\ref{W66}), we obtain the energy $E$ and angular momentum $L$ as follows,
\begin{align}
\label{E2C}
E =&\, \frac{\partial \mathcal{L}}{\partial \dot t} = f(r)\dot{t}=\left(1+\frac{\alpha r^2}{r^3+\beta}\right)\dot{t}\,,\\
\label{L2}
L =&\, \frac{\partial \mathcal{L}}{\partial \dot \phi} = r^2 \dot \phi\,.
\end{align}
Using the above expressions, we obtain the effective potential by rewriting the Lagrangian (\ref{77}),
\begin{align}
\label{17}
\frac{E^2}{1+\frac{\alpha r^2}{r^3+\beta}}-\frac{\dot r^2}{1+\frac{\alpha}{r}+\frac{\beta}{r^3}} - \frac{L^2}{r^2} = \sigma\,.
\end{align}
The corresponding  effective potential of the BH solution (\ref{W66}) takes the following form,
\begin{align}
\label{eq:pot}
\mathcal{V}(r) &= \frac{L^2}{2r^2}\left(1+\frac{\alpha}{r}+\frac{\beta}{r^3}\right)
+ \frac{\sigma}{2} \left(1+\frac{\alpha}{r}+\frac{\beta}{r^3}\right)
 - \frac{E^2\left(1+\frac{\alpha}{r}+\frac{\beta}{r^3}\right)}{2\left(1+\frac{\alpha r^2}{r^3+\beta}\right)} \,.
\end{align}

For circular photon orbits, $\sigma = 0$, solving the zeroth order equations yields,
\begin{align}
\label{ppp1}
0 =-2{r_0}^3{L_0}^2+3{r_0}^2\beta {E_0}^2+6{r_0}^2M {L_0}^2-5\beta {L_0}^2\,,\quad
L_{0\pm} = \pm E_0r_0 \sqrt{\frac{{r_0}^3+\beta}{{{r_0}^3+\beta-2M {r_0}^2}}}\,.
\end{align}
Eq.~(\ref{ppp1}) shows that when the dimensional constant $\beta$ vanishes, we obtain the zeroth order terms of $r_0$ and $L_0$
of the Schwarzschild space-time when $\sigma=0$, i.e., $r_0=3M$ and $ L_{0\pm} =\pm 3\sqrt{3}ME_0$.
The above equation has three roots for $r_0$, one of them has a real value and takes the following form,
\begin{align}
\label{pp1}
r_0=&\, \frac{\sqrt[3]{\beta^3{E_0}^6-10 \beta{L_0}^6+6\beta^2{E_0}^4{L_0}^2M+12 \beta{E_0}^2{L_0}^4M^2-8{L_0}^6M^3-16 S{L_0}^3}}{2{L_0}^2} \nonumber\\
&\, +\frac{ \left( \beta {E_0}^2+2{L_0}^2M \right)^2}{2{L_0}^2 \sqrt[3]{\beta^3{E_0}^6-10 \beta{L_0}^6+6\beta^2{E_0}^4{L_0}^2M
+12 \beta{E_0}^2{L_0}^4M^2-8{L_0}^6M^3-16 S{L_0}^3}}+\frac{\beta\,{E_0}^2+{2ML_0}^2 }{2{L_0}^2}\,,
\end{align}
where $S=\sqrt{5\beta \left( {L_0}^6M^3+5 \beta{L_0}^6-\beta^3{E_0}^6-6\beta^2{E_0}^4{L_0}^2M+12\beta{E_0}^2{L_0}^4M^2 \right) }$.
The above equation gives the value of the Schwarzschild space-time when $\beta=0$.

Now we derive the explicit form of the perihelion shift for massive particles for the BH (\ref{W66}) by using the potential $\mathcal{V}$ (\ref{eq:pot}) with $\sigma=1$.
%, see \eqref{eq:pot}, its first derivative $\mathcal{V}'$ and its second derivative $\mathcal{V}''$.
We evaluate the equations $\mathcal{V}(r_crc.) = 0$ and $\mathcal{V}' \left(r_\mathrm{crc.} \right) = 0$ by considering the perturbation
with $L= L_0 + \epsilon h_1$ and $k = k_0 +\epsilon k_1$.
The zeroth order terms of these equations determine $L_0 \left(r_\mathrm{crc.} \right)$ and $E_0 \left(r_\mathrm{crc.} \right)$ as follows,
\begin{align}
\label{LE3}
E_{0\pm}= \pm \frac{\sqrt{2}({r_\mathrm{crc.}}^3-2M{r_\mathrm{crc.}}^3+\beta)}{\sqrt{2{r_\mathrm{crc.}}^6-6M{r_\mathrm{crc.}}^5+4\beta {r_\mathrm{crc.}}^3+2\beta^2}}\, , \quad
L_{0\pm} = \pm {r_\mathrm{crc.}}^2\sqrt{\frac{2M{r_\mathrm{crc.}}^3+4M\beta}{2r^6-6M{r_\mathrm{crc.}}^5+4\beta {r_\mathrm{crc.}}^3+2\beta^2}}\, .
\end{align}
By using the obtained constants of motion for the circular orbit, we derive the perihelion shift by plugging of the expressios of
$E_0=E_{0\pm}$, $L_0=L_{0\pm}$ into $V'' \left( r_\mathrm{crc.}, E_0, L_0 \right)$.
Corresponding to the signatures $\pm$ in the expressions of $E_0=E_{0\pm}$ and $L_0=L_{0\pm}$, there exist two options to derive the perihelion shift,
\begin{align}
\Delta\phi\left(L_{0+}\right)\,, \quad \Delta\phi\left(L_{0-}\right)\,,
\end{align}
which are related to each other through
\begin{align}
\Delta\phi\left(L_{0-}\right) &= - 4\pi - \Delta\phi\left(L_{0+}\right)\,.
\end{align}
By expanding the perihelion shift into a power series of $q = \frac M{r_\mathrm{crc.}}$ and $q_1 = \frac{\beta}{{r_\mathrm{crc.}}^3}$, we obtain
\begin{align}
\label{eqn:perihelionshift11}
\Delta\phi\left(L_{0-}\right) =2\frac{\sqrt{1-2q_1}-\sqrt{1-6q+11q_1-8q_1{}^2}}{\sqrt{1-6q+11q_1-8q_1{}^2}}\,.
\end{align}
Eq.~(\ref{eqn:perihelionshift11}) when $q_1=0$ yields,
\begin{align}
\label{eqn:perihelionshift2}
\Delta\phi\left(L_{0-}\right) \approx 6 \pi q + 27 \pi q^2-13\pi q_1-105 \pi qq_1 +\frac{435\pi}{4}q_1{}^2 +\mathcal{O}\left( \left( qq_1 \right)^3 \right) \,,
\end{align}
which coincides with the perihelion shift of the Schwarzschild solution when $q_1=0$.

The qualitative behavior of the perihelion shift is not so changed, only the numerical values differ.
As for the photon sphere, the higher $q_1$, the higher the influence of the perturbation and corrections to the perihelion shift appear.

\section{Conclusions}\label{con}

In this study, we constructed a consistent ghost-free modified GB gravitational theory capable of describing BH with horizons.
The field equations of this theory are applied to a spherically symmetric space-time and we succeeded to derive BH solutions with multi-horizons.
We showed that for the Schwarzschild BH type metric (\ref{metric1}), we obtained a BH solution with three horizons
and the curvature invariants of this BH show a true singularity at $r=0$.
%Moreover, we calculated the GB invariant of this BH and showed that it is a dynamic one and always has a positive pattern.
Moreover, we calculated the thermodynamical quantities associated with this solution and showed that all the thermodynamical quantities
and the heat capacity and Gibbs free energy tell that this solution is not stable.

We repeated our calculations for a more general case whose metric is given by (\ref{metric2}) and showed that the solution has two horizons in spite
that the field equations do not include cosmological constant nor there is not any source of charge to reproduce such two horizons.
Moreover, we also showed that such BH yields a true singularity at $r=0$.
We also calculated the thermodynamical quantities and showed that the Gibbs is negative.
Furthermore, for both BH solutions in (\ref{metric1}) and (\ref{metric2}), we calculated all the physical quantities which appear in the GFGB theory,
that is, the potential, the Lagrange multiplier, and the function $f$ and showed their behaviors in FIG.~\ref{Fig:1} and \ref{Fig:3}.

We should note that the present study is a first trial in the direction of a full phenomenological classification of observables, which is derived in the
weak GFGB gravity, to compare them with observations.
The future work in this direction could be to study axially symmetric perturbations around rotating space-time, to obtain the shift in the photon regions,
that will give an important imprint on the predictions of the shape of the BH shadow.
This case will be studied elsewhere.

\end{document}